\documentclass[]{article}

\usepackage{url}
\usepackage{fancyvrb}
\usepackage{graphicx}
\usepackage{booktabs} % For formal tables
\usepackage{tabularx}
\usepackage{makecell}
\usepackage{amsmath}    % Extended typesetting of mathematical expression.
\usepackage{amssymb}    % Provides a multitude of mathematical symbols.
\usepackage{mathtools}  % Further extensions of mathematical typesetting.

\usepackage{subcaption}
\usepackage{authblk}
\usepackage{hyperref}
\usepackage[figuresleft]{rotating}

\title{Maintainable Log Datasets for Evaluation of Intrusion Detection Systems}
\author[1]{Max~Landauer}
\author[1]{Florian~Skopik}
\author[1]{Maximilian Frank}
\author[1]{Wolfgang~Hotwagner}
\author[1]{Markus~Wurzenberger}
\author[2]{Andreas~Rauber}
\affil[1]{Austrian Institute of Technology, Vienna, Austria \authorcr \textit{firstname.lastname@ait.ac.at}}
\affil[2]{Vienna University of Technology, Vienna, Austria \authorcr \textit{rauber@ifs.tuwien.ac.at}}

\begin{document}
	
\maketitle
	
\begin{abstract}
Intrusion detection systems (IDS) monitor system logs and network traffic to recognize malicious activities in computer networks. Evaluating and comparing IDSs with respect to their detection accuracies is thereby essential for their selection in specific use-cases. Despite a great need, hardly any labeled intrusion detection datasets are publicly available. As a consequence, evaluations are often carried out on datasets from real infrastructures, where analysts cannot control system parameters or generate a reliable ground truth, or private datasets that prevent reproducibility of results. As a solution, we present a collection of maintainable log datasets collected in a testbed representing a small enterprise. Thereby, we employ extensive state machines to simulate normal user behavior and inject a multi-step attack. For scalable testbed deployment, we use concepts from model-driven engineering that enable automatic generation and labeling of an arbitrary number of datasets that comprise repetitions of attack executions with variations of parameters. In total, we provide 8 datasets containing 20 distinct types of log files, of which we label 8 files for 10 unique attack steps. We publish the labeled log datasets and code for testbed setup and simulation online as open-source to enable others to reproduce and extend our results.
\end{abstract}
	
\section{Introduction}

Cyber attacks pose a threat to network and system security at any scale. To achieve their goals, which usually range from intrusion, espionage, sabotage, and system takeover, adversaries typically utilize a wide range of tools and attack techniques to discover previously unknown vulnerabilities and find new attack vectors. While system operators seek to keep their network components patched, the ever-changing threat landscape implies that ultimate security is impossible to guarantee as networks continue to grow dynamically over time.

To counteract these problems, manual security-related tasks of system operators have long been supported by automatic tools that continuously monitor networks and systems for both known and unknown threats. Thereby, these so-called intrusion detection systems (IDS) usually ingest network traffic or system log data and analyze their contents for malicious activities. Many IDSs also carry out file integrity checks or scan registry keys and system memories; however, in the context of this paper we solely focus on intrusion detection techniques that leverage log data, i.e., sequentially generated and chronologically ordered events that usually comprise a timestamp and a message containing parameters. IDSs that analyze such log data are most often differentiated into signature-based detection systems, that search for predefined indicators such as hash sums that are known to correspond to malware, and anomaly-based detection systems, that employ self-learning techniques to capture the baseline system behavior and detect any deviations of this learned model as potential threat \cite{khraisat2019survey, chandola2009anomaly}. 

Independent of their type, evaluating IDSs for their ability to detect attacks is crucial to compare different approaches and objectively select appropriate detection techniques for specific system environments. Thereby, publicly available benchmark log datasets are an indispensable prerequisite to enable evaluations. Unfortunately, such log datasets are scarce and usually do not fulfill the requirements set by security researchers. In particular, one of the most crucial aspects of evaluations is to compute detection accuracies, which requires a ground truth that specifies all malicious log events. However, datasets collected from real infrastructures generally lack a reliable ground truth as it is not possible to ensure that only normal and benign activities are carried out on the network, except for purposefully injected attacks \cite{scott1999evaluating}. Moreover, adjusting configurations of components in productive environments or launching attack cases is often only possible in a limited scope since the security and availability of these systems are of utmost importance to the organizations hosting the infrastructures \cite{uetz2021reproducible}. In addition, datasets collected in real environments most often cannot be published due to privacy concerns as log data frequently contains user data or parts of sensitive file contents.

To alleviate problems with real infrastructures altogether, security analysts recreate networks and systems in testbeds and use simulations to generate a base load of normal system operation. However, even datasets created in such controlled environments have been criticized for several reasons, for example, missing documentation that explains installed services \cite{macia2018ugr, al2020real}, limited generalizability \cite{al2020real}, outdated or too simple attack cases \cite{ring2017flow, thomas2008usefulness}, heavy preprocessing such as removal of event parameters \cite{vceponis2018towards}, involvement of closed-source software \cite{thomas2008usefulness, uetz2021reproducible}, lack of periodic behavior \cite{macia2018ugr}, missing reproducibility \cite{uetz2021reproducible}, insufficient duration \cite{creech2013generation}, focus on single hosts rather than the whole network \cite{macia2018ugr}, or lack of variations of attack parameters \cite{scott1999evaluating}.

\begin{figure}
	\centering
	\includegraphics[width=.8\columnwidth]{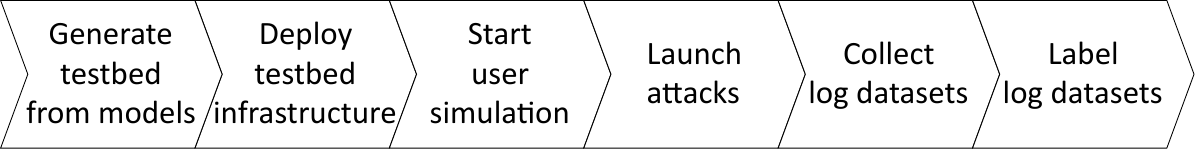}
	\caption{Procedure for generating labeled log datasets.}
	\label{fig:dataset_gen}
\end{figure}

In addition, testbeds generally require a high effort to setup, configure, update, and adjust components. In our earlier work \cite{landauer2020have}, we therefore propose to introduce concepts from model-driven engineering in testbed deployment processes. Figure \ref{fig:dataset_gen} visualizes our procedure for generating labeled log datasets from model-driven testbeds. Contrary to common testbed generation approaches that result in single static test environments, our approach implies to generate models for infrastructure setup, normal behavior simulation as well as attack execution that act as templates by leaving out several parameters as variables, and to define transformation rules that dynamically fill out these parameters when launching a testbed. The main advantage of this methodology is that it is simple to generate an arbitrary number of datasets that stem from different testbeds with variations, i.e., normal and malicious traces are slightly different across datasets and thus enable more robust evaluations. We recognize some shortcomings of our implementation, including a fairly simple network structure and an unreliable labeling strategy. To overcome these problems, we largely extend the scope of our simulation and integrate an automatic labeling mechanism \cite{landauer2020kyoushi}.

Alongside with this paper, we publish a collection of log datasets generated with the presented approach as well as all code that is necessary to run our testbed and simulations within it so that other researchers are able to replay or augment the simulation runs. Our datasets are therefore maintainable and allow for continuous improvements such as enlargements of the labeling range as well as additions of datasets from new testbeds. We summarize our contributions as follows:

\begin{itemize}
	\item A publicly available labeled collection of log datasets for evaluation of IDSs\footnote{AIT-LDSv2.0 \cite{aitlds2} available at \url{https://zenodo.org/record/5789063}},
	\item an analysis and comparison of these datasets with respect to real user logs,
	\item an open-source implementation to launch testbeds for dataset generation\footnote{Kyoushi testbed environment, \url{https://github.com/ait-aecid/kyoushi-environment}}, and
	\item an open-source library\footnote{Kyoushi simulation package, \url{https://github.com/ait-aecid/kyoushi-simulation}} and models\footnote{Kyoushi simulation models, \url{https://github.com/ait-aecid/kyoushi-statemachines}} to simulate normal user behavior and attacker activities.
\end{itemize}

The remainder of this paper is structured as follows. Section \ref{related} reviews existing log datasets. In Sect. \ref{methodology} we outline our methodology for generating log datasets and explain our modeled scenario. We analyze the generated datasets in Sect. \ref{analysis} and discuss the results in Sect. \ref{discussion}. Finally, Sect. \ref{conclusion} concludes the paper.

\section{Background and Related Work} \label{related}

Due to the large need for datasets in cyber security research, several attempts to generate benchmark datasets were made in the past. However, most of these datasets are created with specific use-cases in mind and are thus not generally applicable. To compare these datasets on a common scheme, we first describe a set of requirements that are relevant for intrusion detection datasets and then discuss the fulfillment of these aspects for several state-of-the-art datasets.

\subsection{Requirements} \label{requirements}

Recording log datasets in testbeds or real environments is not straightforward; it is a task that requires careful planning, since the quality and usefulness of the resulting data strongly relies on several decisions made by the analyst. We gathered a list of requirements by reviewing design principles that were followed by authors of existing datasets. In the following, we summarize our findings.

\begin{enumerate}
	\item \textbf{Use-case}. To ensure relevance and authenticity of the dataset, it is necessary to design the overall network layout and technical infrastructure of the system where log data is recorded in the context of a specific scenario. This also includes services available on the involved machines \cite{landauer2020have}. Clearly specifying the scope of the simulation also helps to define the limitations of the dataset.
	\item \textbf{Synthetic data generation}. Datasets collected from real-world system environments are sometimes considered superior to synthetically generated data due to the fact that they are per definition realistic, while simulations only try to replicate their characteristics \cite{haider2017generating}. However, real datasets have the strong disadvantage that it is infeasible to differentiate normal from anomalous or malicious logs with complete certainty, since the root causes of some actions are unknown to the analysts \cite{scott1999evaluating}. Obviously, synthetic dataset generation implies that scripts that replicate normal behavior on an appropriate level of detail are prepared beforehand. This particularly concerns models for user activities that normally occur on the system, which can be very diverse and thus non-trivial to formalize. On the plus side, modeling the normal behavior effectively enables to steer the parameters of the simulation to generate data that is representative for different levels of detection complexity  \cite{scott1999evaluating}. Therefore, we argue that synthetically generated log datasets are the best option for IDS evaluations.
	\item \textbf{Attacks}. As part of a realistic evaluation of IDSs, it is necessary to select recent and relevant attack scenarios that are suitable for the system environment at hand \cite{ring2017flow, landauer2020have}. Otherwise, outdated attack cases may not yield representative intrusion detection evaluation results that are comparable to that of more modern attacks.
	\item \textbf{System logs}. When IDSs are applied in productive systems, they are usually able to analyze logs in raw and unaltered form. Accordingly, log datasets for evaluation of IDSs should also provide logs that are not processed in any way \cite{macia2018ugr}. Fortunately, synthetic datasets recorded in simulations are usually less critical when it comes to privacy, since no humans are involved and thus anonymization of personal user data that possibly occurs in the logs is not required. This also concerns sensible contents of files that may appear in the logs and should thus be simulated with collections of predefined dummy files \cite{uetz2021reproducible}. Another important aspect is to configure the logging framework in a realistic way that fits the use case. For this, analysts must decide where to log and what to log \cite{zhu2015learning}. In particular, anomaly-based IDSs require logs corresponding to normal system behavior to learn a baseline for detection, meaning that logging levels should be set to \textit{info} or even \textit{debug} rather than \textit{error} or \textit{warning}. Moreover, it is beneficial to log performance metrics such as CPU or memory data, because they are also adequate inputs for IDSs \cite{khorshed2011monitoring}.
	\item \textbf{Network traffic}. Beside system logs that are the main input of host-based IDSs, network traffic is a widely used data source for network-based IDSs. Accordingly, datasets should also include packet captures to enable evaluation of network-based IDSs and hybrid IDSs that make use of both system logs and network traffic \cite{liao2013intrusion}.
	\item \textbf{Periodicity}. Productive system environments naturally exhibit periodic behavior, for example, cron jobs are scheduled for execution in fixed intervals and events originating from human activities follow daily and weekly patterns of work shifts. Self-learning IDSs are able to integrate these cycles in their models to detect contextual anomalies, i.e., events that are considered anomalous due to their time of occurrence \cite{chandola2009anomaly}. It is therefore essential to expand the duration of the simulation to cover several of these cycles \cite{macia2018ugr}.
	\item \textbf{Labels}. Ground truth tables that unambiguously assign labels to all events are needed to compute evaluation metrics such as detection accuracy or false alarm rates \cite{ring2017flow}. Accordingly, it is essential to provide a comprehensible methodology for creating correct ground truth tables for IDS evaluation.
	\item \textbf{Documentation}. Datasets should be published with detailed descriptions of all relevant aspects of the data creation. Otherwise, it is not possible for others to fully understand all artifacts present in the data, which could possibly lead to incorrect assumptions and invalidate evaluation results \cite{macia2018ugr}.
	\item \textbf{Repetitions}. For anomaly-based IDSs that only learn from normal behavior and then classify test data either as normal or anomalous, it is sufficient to only have artifacts of a single attack execution in the data. However, for attack classification it is necessary that attacks are at least present in training and test datasets, and possibly validation datasets. Accordingly, attacks should be launched multiple times by repeating the simulation. In addition, research on alert aggregation urgently requires useful datasets, especially for system logs analyzed by host-based IDSs \cite{navarro2018systematic}. Thereby, clustering-based aggregation methods require that the same attacks are carried out multiple times to form groups \cite{landauer2020dealing}.
	\item \textbf{Variations}. Approaches for both attack classification and alert aggregation should be challenged by introducing variations in attack executions \cite{scott1999evaluating}. Moreover, evaluation results have a higher robustness when they are based on multiple attack executions that cover a spectrum of possible attack variations \cite{landauer2020have}. This behavior could be realized by dynamically changing attack parameters in each simulation run.
	\item \textbf{Reproducibility}. Technologies that constitute the simulation are continuously updated. To avoid that datasets become outdated, it should be possible to repeat simulations at any given time \cite{thomas2008usefulness, uetz2021reproducible}. This also allows to reuse existing assets and only change certain parts of the simulation, e.g., keep the infrastructure and user simulation, but include another attack vector. It is therefore beneficial to publish all code used to carry out the simulation alongside the resulting datasets.
\end{enumerate}

\subsection{Literature Analysis} \label{literature}

The previous section outlines a set of requirements that should be fulfilled by datasets to enable evaluation of intrusion detection systems. We gathered several datasets that are commonly used in scientific evaluations and analyzed whether they fulfill our requirements. Table \ref{tab:sota} shows a complete list of all datasets and our findings, where \checkmark indicates that the datasets fulfill the respective requirement, $\sim$ indicates partial fulfillment, and no symbol means that the requirement is not fulfilled. In the following, we discuss our findings and relevant properties of the datasets in detail.

\begin{sidewaystable*}[]
	\centering
	\footnotesize
	\caption{Fulfillment of requirements for existing datasets}
	\begin{tabular}[t]{lccccccccccc}
		& \multicolumn{11}{c}{\textbf{Requirement}} \\ \hline
		\textbf{Dataset} & \textbf{(1)} & \textbf{(2)} & \textbf{(3)} & \textbf{(4)} & \textbf{(5)} & \textbf{(6)} & \textbf{(7)} & \textbf{(8)} & \textbf{(9)} & \textbf{(10)} & \textbf{(11)} \\ \hline \hline
		ADFA-LD \cite{creech2013generation} & Linux OS & & \checkmark & $\sim$ & & & \checkmark & \checkmark & \checkmark & & \\ \hline
		ADFA-WD \cite{creech2014developing} & Windows OS & & \checkmark & $\sim$ & & & \checkmark & \checkmark & \checkmark & & \\ \hline
		IoT-DDoS \cite{al2020real} & Internet of Things & \checkmark & \checkmark & & \checkmark & $\sim$ & \checkmark & & & & \\ \hline
		AWSCTD \cite{vceponis2018towards} & Windows OS & & \checkmark & \checkmark & \checkmark & $\sim$ & & \checkmark & & & \\ \hline
		CIDD \cite{kholidy2012cidd} & Cloud Systems & & \checkmark & \checkmark & \checkmark & \checkmark & \checkmark & \checkmark & & & \\ \hline
		CIDDS \cite{ring2017flow} & Enterprise IT & \checkmark & \checkmark & & \checkmark & \checkmark & \checkmark & & & & \\ \hline
		LID-DS \cite{grimmer2019modern} & Linux OS & \checkmark & \checkmark & \checkmark & & & \checkmark & \checkmark & \checkmark & & \\ \hline
		VAST Challenge 2011 \cite{vast_2011} & Enterprise IT & & \checkmark & \checkmark & \checkmark & \checkmark & \checkmark & \checkmark & & & \\ \hline
		KDD Cup 1999 \cite{kdd99} & Military IT & & & & \checkmark & \checkmark & \checkmark & \checkmark & \checkmark & \checkmark & \\ \hline
		Loghub \cite{he2020loghub} & Supercomputer and OS & & & \checkmark & & \checkmark & & \checkmark & & & \\ \hline
		NGIDS DS \cite{haider2017generating} & Enterprise IT & \checkmark & \checkmark & & \checkmark & & \checkmark & & & & \\ \hline
		CICIDS 2017 \cite{sharafaldin2018toward} & Enterprise IT & \checkmark & \checkmark & & \checkmark & \checkmark & \checkmark & \checkmark & & & \\ \hline
		Skopik et al. \cite{skopik2014semi} & Enterprise IT & \checkmark & \checkmark & \checkmark & \checkmark & & & \checkmark & & & \\ \hline
		SOCBED dataset \cite{uetz2021reproducible} & Enterprise IT & \checkmark & \checkmark & \checkmark & $\sim$ & & & \checkmark & \checkmark & $\sim$ & \checkmark \\ \hline
		UGR'16 \cite{macia2018ugr} & Enterprise IT & & \checkmark & & \checkmark & \checkmark & \checkmark & \checkmark & \checkmark & & \\ \hline
		HDFS \cite{xu2009detecting} & Supercomputer & & & \checkmark & & $\sim$ & \checkmark & & & & \\ \hline
		AIT-LDSv1.1 \cite{landauer2020have} & Enterprise IT & \checkmark & \checkmark & \checkmark & & \checkmark & $\sim$ & \checkmark & \checkmark & \checkmark & \\ \hline
		\textbf{AIT-LDSv2.0} (this paper) & Enterprise IT & \checkmark & \checkmark & \checkmark & \checkmark & \checkmark & \checkmark & \checkmark & \checkmark & \checkmark & \checkmark \\ \hline
	\end{tabular}
	\label{tab:sota}
\end{sidewaystable*}

One of the earliest log datasets that became widely used in intrusion detection is the KDD Cup 1999 dataset \cite{kdd99}. The logs were collected during a simulation of several intrusions in a military network. Other than many modern datasets, the authors made sure to label all events with the respective attack types and furthermore repeat and vary the attacks to yield different probability distributions in the training, validation, and test datasets. These properties make this dataset especially attractive for evaluating machine learning techniques. Even today it is still widely used in scientific publications, although the dataset has been repeatedly criticized for being outdated, too simple, and not reproducible due to the fact that closed-source tools were used for traffic generation \cite{thomas2008usefulness}. 

As a consequence of these criticisms, Creech et al. generated ADFA-LD \cite{creech2013generation} and ADFA-WD \cite{creech2014developing}, two datasets containing sequences of system calls on a Linux and Windows host respectively. For the generation of the dataset, the authors simulated normal activities such as web browsing and file editing and launched several attacks, such as brute-force logins and exploits for webshell uploads. Unfortunately, the system calls are stripped from all contextual variables such as timestamps, parameters, and return values, and are thus not representative for real data \cite{vceponis2018towards}. Moreover, the dataset is criticized for only including a single host, not generalizing well for other systems, as well as a lack of documentation detailing how the dataset is collected and what services are installed \cite{al2020real, macia2018ugr}. The AWSCTD \cite{vceponis2018towards} aims to resolve at least one of these issues by recording Windows system calls without removing any parameters and further extend the set of launched attacks. However, the authors also consider only a single host and not a full network.

Another dataset based on Linux system calls is LID-DS \cite{grimmer2019modern}. While the authors explain the attack scenarios in great detail, there is only little information on the simulation of normal system behavior. They carry out all attacks multiple times and collect the logs from hundreds of runs that last around 30 seconds each. CIDD \cite{kholidy2012cidd} provides logs specifically for masquerade attacks. One of the noteworthy aspects of this data is that the authors manage to label all events by correlating network and system logs and mapping them to attack tables specifying the expected times, IP addresses, and user names related to attacks. Moreover, the users generating normal activity in the dataset are categorized into normal, advanced, administrators, programmers, and secretary users.

One of the few datasets that also include system logs other than system calls is from the VAST Challenge 2011 \cite{vast_2011}. In particular, the dataset comprises firewall logs, IDS alerts, syslogs and network packet captures. Among the attacks launched against the simulated system are security scans, denial-of-service attacks, and remote desktop connections as consequences of a social-engineering attack. The authors also provide a document describing the solutions to the challenge that depict a ground truth of malicious events. The dataset presented alongside the open-source testbed called SOCBED \cite{uetz2021reproducible} contains system logs from a network of Windows and Linux hosts. While the authors did not collect network traffic for this dataset, they state that it is simple to extend their testbed accordingly and repeat the experiments. In addition, the authors discuss variations in log data, however, only with respect to circumstantial factors such as system performance and not purposefully incorporated variations as accomplished by our model-driven approach. Skopik et al. \cite{skopik2014semi} also collect network traffic as well as access and application logs on a testbed where simulated users click around on a mail platform. Contrary to most existing papers that present new datasets, they configure their user simulations based on behavior of real users and also validate their data by comparison of accessed resources. Other datasets comprising system and application logs from various services are provided in Loghub \cite{he2020loghub}. The main problem with these datasets is that they mainly involve anomalous traces related to failures rather than cyber attacks.

While system log datasets are most often collected from single hosts, whole networks comprising several hosts are usually deployed to generate network traffic datasets. For generating CIDDS \cite{ring2017flow}, the authors recreated a virtual company with network components that are commonly used in enterprise IT, e.g., Windows and Linux hosts as well as file shares and web servers, and place them in separate subnets for managements, office, and developers. Their user simulations are based on state machines to generate complex behavior patterns instead of repeated sequences and their models also respect working hours and breaks. Moreover, their network is also connected to the Internet to mix the simulated traffic with real connections and possibly attacks. To generate the UGR'16 \cite{macia2018ugr}, the authors also use a combination of real user behavior traffic and simulated attack traffic. Thereby, the authors specifically pay attention to the cyclic behavior of communication logs that originates from daily or weekly usage patterns. Moreover, their attacks are generated with random starting times.

The authors of CICIDS 2017 \cite{sharafaldin2018toward} follow a different approach as make use of a profiler that analyzes real communication in a network and then arbitrarily generates data following these patterns. They recorded the network traffic while launching several attacks, among which are denial-of-service attacks, vulnerability exploits, and a botnet. Similarly, a network traffic generation appliance was also used to generate NGIDS DS \cite{haider2017generating}. Other than these datasets, the IoT-DDoS \cite{al2020real} specifically focuses on a scenario that simulates Internet of Things in a network.

In our earlier work we present the AIT-LDSv1.1 \cite{landauer2020have}, a system log dataset collected from a webserver hosting a content management system and groupware. Other than most existing approaches for dataset generation, the paper \cite{landauer2020have} describes a model-driven strategy for automatic testbed deployment to generate multiple datasets with variations of attack executions. We recognize several shortcomings of the dataset: First, beside some machines running user simulations, the network is relatively simple as it only consists of a single webserver. Second, the simulation focuses on system log data and thus no network traffic is captured. Third, labeling of malicious events is not reliable since it relies on similarity-based matching, which may lead to incorrectly unlabeled lines in case that variations lead to new or dissimilar events \cite{landauer2020kyoushi}. Finally, only the resulting data is publicly available, but the scripts for deploying the testbed and running the simulation are not accessible. As a consequence of these shortcomings, we propose the AIT-LDSv2.0 in this paper. In comparison to our previous testbed for log data generation, we increased the network complexity, collected logs from all components of the network (e.g., the firewall), extended the simulation of normal behavior, improved the strategy for event labeling, and published all code for deploying the testbed along with the generated dataset. As visible in Table \ref{tab:sota}, our new dataset meets all requirements stated in Sect. \ref{requirements}. We discuss the fulfillment of these requirements in detail in Sect. \ref{fulfillment}.

\section{Log Dataset Generation Methodology} \label{methodology}

This section outlines the overall methodology for the generation of our dataset. We first describe a procedure for automatic testbed deployment that leverages concepts from model-driven engineering to enable the generation of multiple datasets with variations. Subsequently, we explain the application scenario modeled by our testbed and state relevant design criteria for the monitored network, simulations of normal behavior, and injected attacks.

\subsection{Testbed Generation} \label{testbed_gen}

In our earlier work \cite{landauer2020have} we presented a model-driven methodology for testbed generation. We also published a follow-up paper that proposes a strategy for log event labeling \cite{landauer2020kyoushi}. In this paper, we combine both methods to generate multiple labeled datasets with variations. In the following, we briefly summarize the main aspects of model-driven testbeds and the integration of our labeling procedure.

As described in Sect. \ref{related}, synthetic log datasets are commonly collected on testbeds, i.e., one or more virtual machines deployed in isolated networks. Thereby, setting up such testbeds involves time- and resource-consuming tasks that often require a high amount of domain knowledge. The resulting testbeds are often relatively static, i.e., difficult to modify in hindsight when updates of certain components are required or changes of the scenario become necessary \cite{landauer2020have}. Model-driven testbed generation, however, alleviates these problems by allowing analysts to design testbeds on a higher level of abstraction and use these models to automatically instantiate arbitrary numbers of testbeds. On top of that, it is possible to specify parameter spaces rather than specific values for all kinds of testbed properties, including network size, frequencies of user interactions, or attack attributes. Datasets generated from such approaches include variations of the system environment as well as user and attacker behavior that manifest in the logs, which is beneficial for IDS evaluation since it increases robustness of the results and makes the datasets applicable for evaluation of alert aggregation.

\begin{figure}
	\centering
	\includegraphics[width=.8\columnwidth]{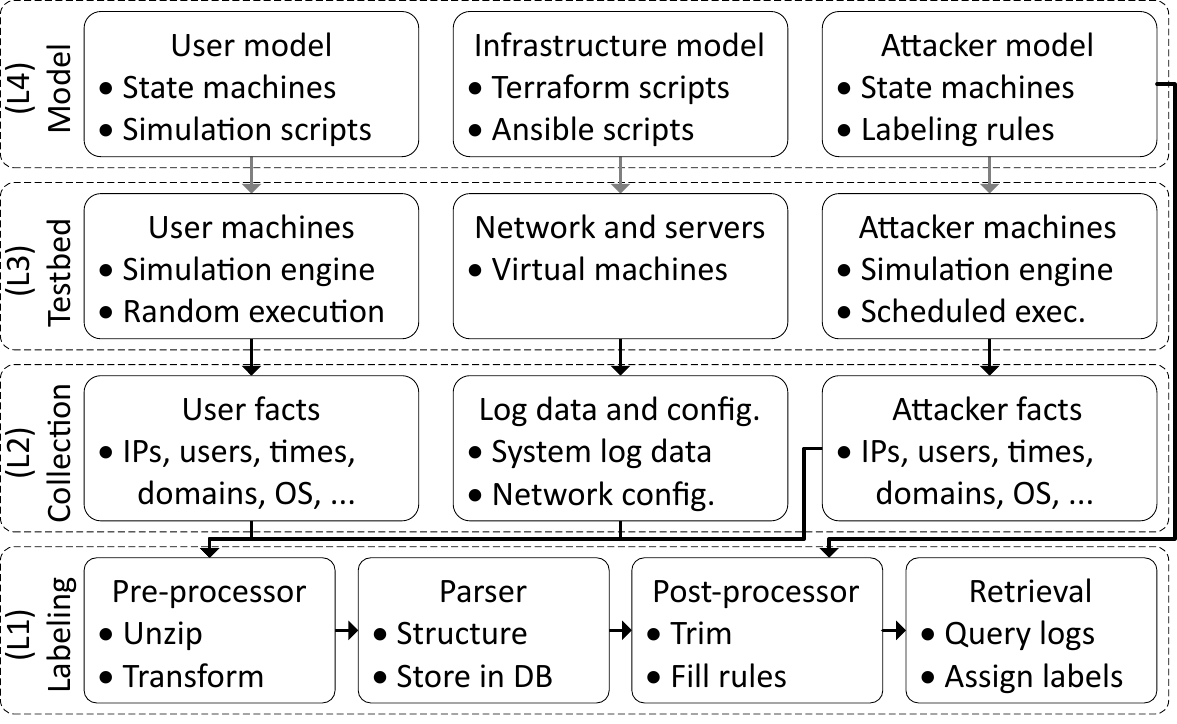}
	\caption{Concept for model-driven testbed generation and dataset labeling.}
	\label{fig:concept}
\end{figure}

Figure \ref{fig:concept} depicts an overview of the layers involved in such a model-driven approach for dataset generation. Layer (L4) represents the highest level of abstraction and comprises three different types of models: (i) state machines and scripts that simulate normal user behavior, (ii) provisioning scripts for deploying and configuring the technical infrastructure, and (iii) scripts that launch attacks and rules that assign labels to the generated events. All of these models are designed as templates, i.e., they leave out several parameters that are dynamically filled out when instantiating a specific testbed based on predefined ranges and lists. For example, IP addresses of all components are randomly chosen from pools, user names are selected from databases, and transition probabilities are calculated from predefined distributions. Accordingly, we refer to the templated scripts on layer (L4) as testbed-independent models (TIM).

Layer (L3) contains so-called testbed-specific models (TSM) that are instantiated from the TIMs. This is accomplished by running a transformation engine that processes all templates provided by the TIMs and fills out all parameters according to their predefined spaces. Note that this process is fully automatic and can be repeated as often as needed to generate any desired number of testbeds. The resulting TSMs are runnable scripts that are ready to be executed in order to deploy the virtual machines, configure all services, start the user simulations, and launch the attacks at a given point in time. The simulation then runs in real-time to ensure that all generated log artifacts, e.g., timestamps and latency times, resemble execution in real-world scenarios.

Once the simulation is completed, i.e., the analyst determines that logs from a sufficiently long time period are collected or a predefined end time for the simulation is reached, layer (L2) handles the collection of log data from all machines. This mainly involves logs that are typically analyzed by IDSs, e.g., access logs, authentication logs, monitoring logs, and audit logs, but also custom logs generated by our state machines that simulate normal user and attacker behavior. In addition, the collection script gathers so-called facts from all machines, including their IP addresses, OS information, network configurations, etc. These data are necessary for the automatic generation of a ground truth, which is carried out on layer (L4). Labeling consists of a sequence of steps \cite{landauer2020kyoushi}. First, a pre-processor prepares all logs for the following tasks. This includes unzipping archived log files or transforming logs from binary format into text. Second, a parser runs over all log lines, transforms them into tokens, and loads them into a database so that it is possible to query single or multiple logs based on their event parameters. Third, a post-processor trims all stored logs according to the predefined start and stop times of the simulation. Moreover, all labeling rules that are defined as templates within the attacker TIMs are filled out using the facts collected in layer (L2). For example, a rule that labels all DNS log events involving the domain address of the attacker may be automatically augmented with this information by extracting the address as a fact from the attacker's host machine. Similarly, start and stop times that are retrieved as facts from the attack execution logs may be used to limit the search scope of the queries. In the final step of the labeling procedure, the completed labeling rules are used to query logs from the database and assign labels to the results. The main advantage of leveraging facts in rules is that no manual adjustments need to be made when executing the same rules on other testbeds, since all relevant information is automatically extracted as facts from the respective hosts. Finally, after all labeling rules are processed, the resulting dataset consisting of the raw logs and their assigned labels is ready to be shared or used for IDS evaluation.

\subsection{Scenario}

The previous section outlined a general overview of the methodology for the generation of our dataset. In this section, we describe our targeted use-case and explain specific design decisions regarding variations in the dataset.

\subsubsection{Use-case} \label{usecase}

The purpose of our collection of log datasets is to enable evaluation of IDSs in the context of a widespread application scenario that is frequently subject of cyber attacks. Specifically small- or medium-sized organizations are a frequent target of cyber attacks, often due to the fact that they do not have the required resources for extensive protection \cite{symantec}. We therefore design our testbed to resemble a small enterprise network that follows well-known security guidelines, such as segmentation of networks into zones \cite{ISO27033}. 

\begin{figure}
	\centering
	\includegraphics[width=.8\columnwidth]{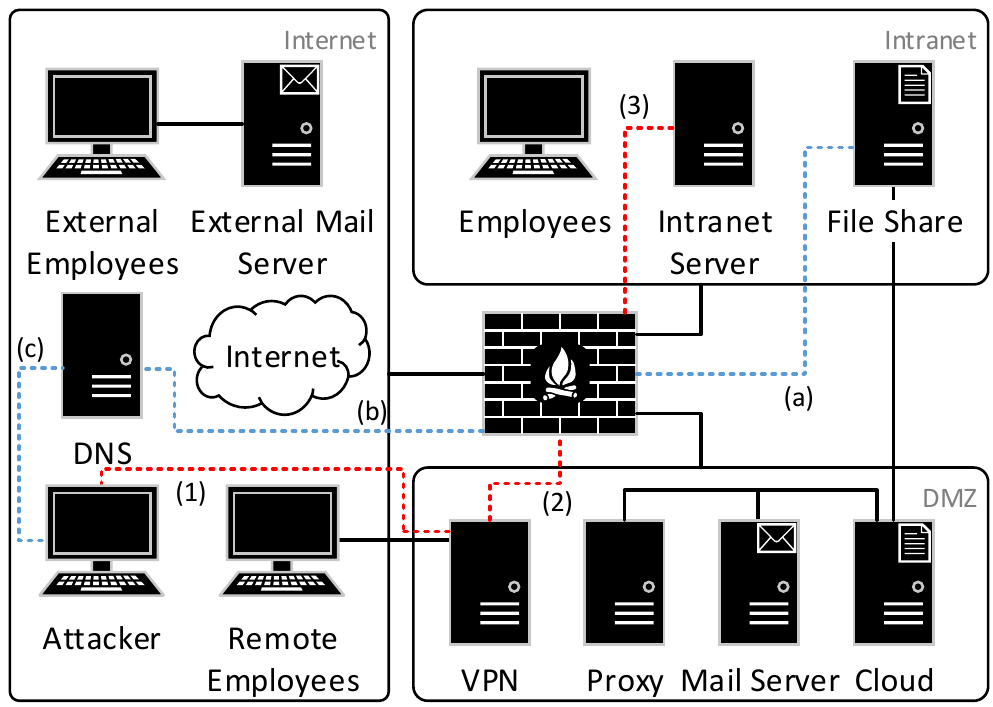}
	\caption{Overview of the testbed network. Steps (1)-(3) mark the attacker's path to compromise the intranet server and steps (a)-(c) represent connections related to the data exfiltration attack vector.}
	\label{fig:network}
\end{figure}

Figure \ref{fig:network} displays an overview of the network realized by our testbed. The network comprises three zones: (i) the intranet that contains a number of Linux hosts\footnote{Ubuntu 20.04, \url{https://ubuntu.com/}} for each employee as well as an intranet server running WordPress\footnote{Wordpress 5.8.2, \url{https://wordpress.com/}} and Samba file share\footnote{Samba 4.5.9, \url{https://samba.org/}}, (ii) the demilitarized zone (DMZ) that contains servers for VPN\footnote{OpenVPN 2.4.4, \url{https://openvpn.net/}}, proxy, mail\footnote{Horde Groupware 5.2.17, \url{https://horde.org/apps/webmail}}, and cloud share\footnote{OwnCloud 10.5.0, \url{https://owncloud.com/}}, and (iii) the Internet with global DNS\footnote{MaraDNS 2.0.13, \url{https://maradns.samiam.org/}, and Dnsmasq 2.79, \url{https://thekelleys.org.uk/dnsmasq/doc.html}}, hosts for remote employees that connect to the intranet via VPN, external employees that use external mail servers, and an attacker host. The zones are connected via a firewall\footnote{Shorewall 5.1.12.2, \url{https://shorewall.org/}} that also acts as an internal DNS server for all domains owned by the organization. All employed technologies are publicly available and commonly used in real networks \cite{macia2018ugr}. 

As outlined in Sect. \ref{testbed_gen}, TIMs result in different TSMs due to the fact that several parameters are set dynamically during instantiation of the testbeds. With respect to the system environment, this mainly concerns the network size and allocation of IP addresses. In particular, we generate between 3 and 9 hosts for internal, remote, and external employees respectively, meaning that the final testbed may consist of at least 9 and at most 27 user simulations running in parallel. Similarly, we generate between 2 and 4 external mail servers. We also assign each network zone a random class and randomly choose IP addresses from these zones for each host. Finally, we also configure the domain names of all network zones as random names using the Faker library\footnote{Faker, \url{https://github.com/joke2k/faker}}. Table \ref{tab:infrastruct} provides a summary of all variations of the technical infrastructure.

\begin{table}[]
	\centering
	\footnotesize
	\caption{Variations of the system environment}
	\begin{tabular}[t]{ll} \hline
		\textbf{Parameter} & \textbf{Range} \\ \hline \hline
		Number of user hosts & 9-27 \\ \hline
		Number of mail servers & 2-4 \\ \hline
		Network zone classes & {[ a, b, c ]} \\ \hline
		Host IPs & Random IP within respective zones \\ \hline
		Network and zone names & Random names \\ \hline
	\end{tabular}
	\label{tab:infrastruct}
\end{table}

\subsubsection{User simulation} \label{usersim}

Real networks in small- or medium-sized organizations are actively used by humans that carry out their daily routines in their workplace. The simulation of normal behavior is therefore an essential aspect of synthetic dataset generation for IDS evaluation. Simulated normal system behavior that is not sufficiently complex may result in non-representative datasets that yield too low false positive rates during IDS evaluation, as human interactions with machines are often erratic and possibly lead to unexpected system states that may be incorrectly detected as malicious. We therefore decided to create state machines for all services in our testbed that are normally accessed by real users. For this purpose, we make use of web automation software\footnote{Selenium, \url{https://www.selenium.dev/}} that allows to use scripts to navigate on websites and click on specific links.

\begin{figure*}
	% left to right, dot, desugar, smcat 9.2.0
	\centering
	\includegraphics[width=1\textwidth]{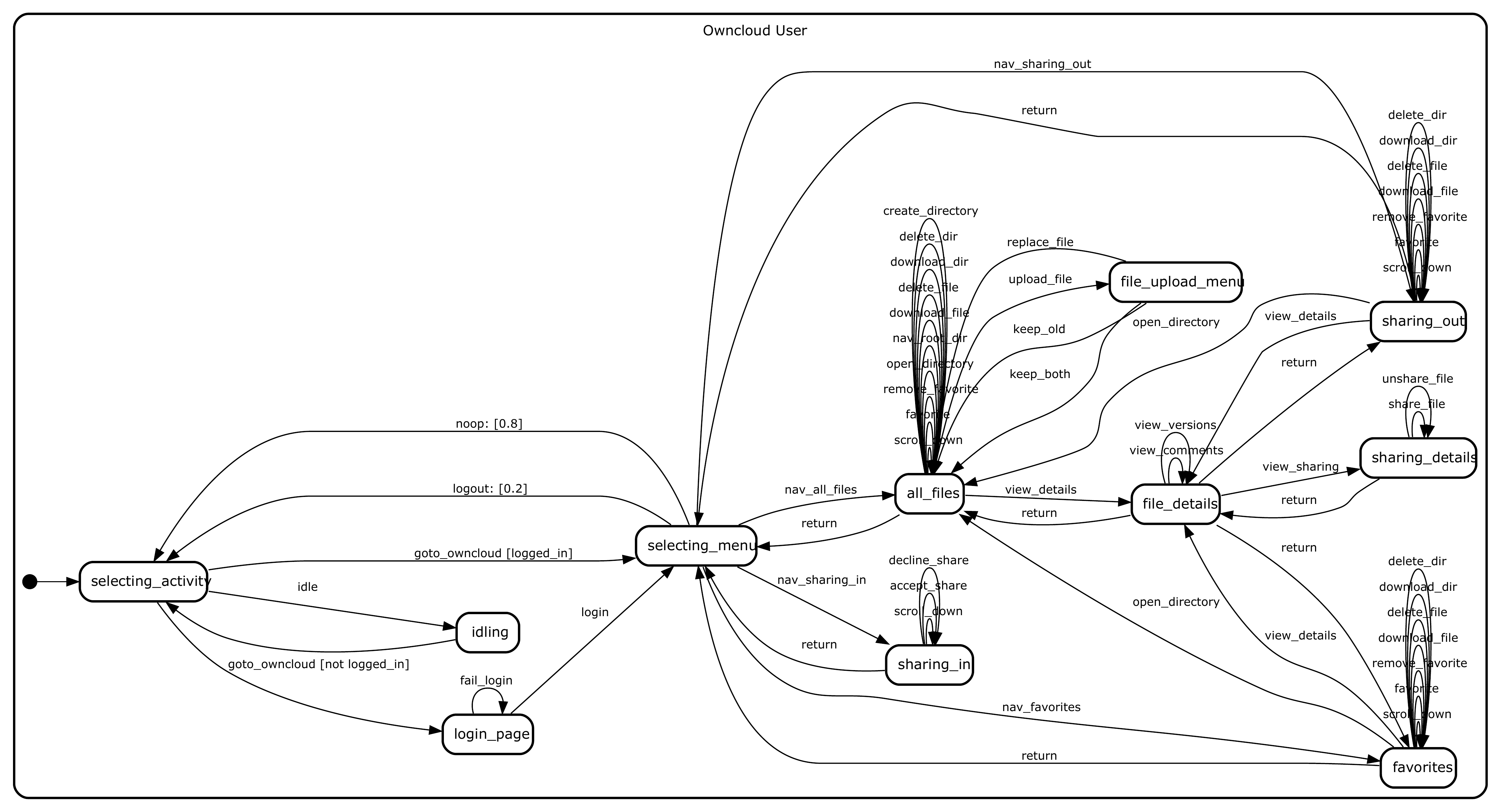}
	\caption{User state machine for simulating normal behavior on the cloud share platform.}
	\label{fig:owncloud}
\end{figure*}

Figure \ref{fig:owncloud} visualizes the state machine for a user accessing the cloud share platform. Note that states describe the current view of the users and that activities such as clicking buttons are carried out when traversing from one state to another. As visible in the figure, the user first logs into the OwnCloud platform (possibly with incorrect credentials, in which case login is retried) and then enters pages showing either all their files, files marked as favorites, files shared with other users, or files other users shared with them. Depending on their selection, the users are then able to view files, upload and share new files, change or remove existing shares, accept or decline invitations to share files, and manage their favorites. Furthermore, there is the possibility that a user leaves the cloud sharing application and switches to another website, or enters the idle state in which case no action is carried out for a certain amount of time. We argue that the total number of possible transitions and interweaving of states visible in Fig. \ref{fig:owncloud} is sufficiently complex to represent real user interaction. Section \ref{normal} will compare log data generated by simulated and real users to verify this claim.

We do not provide figures for all state machines for brevity, but briefly discuss their main features. (i) The web mail state machine allows users to view, compose, and respond to mails from other users, attach files to mails, change their preferences, and manage their calendar entries, contacts, notes, and tasks. In addition, privileged users may access the administrator panel to view and change settings of the platform. (ii) The WordPress state machine allows users to read existing posts on the WordPress instance, publish new posts, comment on existing posts, and view available media. (iii) The Internet state machine allows users to browse the Internet by randomly clicking on links on one of the websites from a predefined list. (iv) The SSH state machine allows users to connect to a host in the network via SSH to execute some of a predefined list of commands. All state machines are connected with each other, i.e., users are able to change between state machines, to further increase the complexity of the simulation.

\begin{table}[]
	\centering
	\footnotesize
	\caption{Variations of simulated user behavior}
	\begin{tabular}[t]{ll} \hline
		\textbf{Parameter} & \textbf{Range} \\ \hline \hline
		User name & Random name \\ \hline
		Password & Random string \\ \hline
		Wordpress role & {[ editor, admin, none ]} \\ \hline
		SSH admin & {[ yes, no ]} \\ \hline
		Samba role & {[ employee, mgmt., acc., admin, none ]} \\ \hline
		OwnCloud role & {[ employee, mgmt., acc., admin, none ]} \\ \hline
		Working hours & (5:00-9:00) - (17:00-22:00) \\ \hline
		User mail provider & Random selection from all mail servers \\ \hline
		User mail contacts & Random selection from all users \\ \hline
		\makecell[lt]{State transition \\ probabilities} & 0.0-1.0 \\ \hline
		Web browser & {[ firefox, chromium ]} \\ \hline
		Idle times & \makecell[lt]{Tiny: 0.4-2.5 seconds \\ Small: 3-60 seconds \\ Medium: 40-360 seconds \\ Large: 400-3600 seconds} \\ \hline
	\end{tabular}
	\label{tab:users}
\end{table}

Whether a user accesses specific states within the state machines or not depends on their roles, which are subject to variation. In particular, we define an SSH administrator role and furthermore differentiate between editor and administrator roles on the WordPress page and employee, management, accounting, administrator roles on Samba and OwnCloud pages. When no role is assigned to a user, the respective state machine is not entered at all. The names of all users are randomly generated from databases and their passwords are random strings. We also vary their working hours, assign their preferred web browser, generate their mail addresses from one of the external mail servers, and select random samples for their usual contacts and available files. To ensure that all files involved in the simulation appear realistic and do not only involve completely randomized contents, we make use of a collection of predefined dummy files with non-sensitive contents. Table \ref{tab:users} provides an overview of the varied parameters and their parameter spaces. Note that we use idle times to temporarily pause the state machines not only in idle states that are specifically created for this task, but also when entering or leaving certain states. This accomplishes to simulate delays between single clicks (tiny), pauses for reading and reacting to website contents (small and medium), or longer breaks of inactivity (large). The table leaves out several minor parameters, such as limits for maximum daily accesses or factors that make repeated executions of same activities more unlikely, for which we refer to our open-source implementation.

\subsubsection{Attack scenario} \label{att_sce}

While simulation of normal user activity is necessary to ensure authenticity of the underlying conditions, injected attacks are required to provide the artifacts to be detected or classified by IDSs. Accordingly, it is essential to design relevant attack cases that fit the overall use-case and are suitable to generate desired consequences in the dataset. For our use-case, we decided to model a multi-step attack that involves several stages of a typical cyber kill chain \cite{fireeyeapt} and makes use of common penetration testing tools \cite{kali}. Figure \ref{fig:network} shows the connections and affected hosts of this attack scenario. In particular, steps (1)-(3) show how the attacker first accesses the intranet over VPN to gather information and eventually takeover the intranet server, and steps (a)-(c) indicate how data is extracted from the file share in the intranet zone over a public DNS server to the attacker. In the following, we explain all attack steps in detail.

As part of our attack scenario, we assume that the attacker illegitimately obtained VPN credentials that allow them to access the network. In real-world attack cases, obtaining such credentials could be achieved through phishing attacks or by compromising a personal computer of an employee. Note that we do not simulate this part of the multi-step attack, since it occurs outside of the enterprise's network and thus does not leave any traces in the logs. 

Once the attack execution starts, the attacker makes use of the VPN credentials to remotely establish a connection to the network over the VPN server. The first step of the attack chain then consists of several scans of the network. In particular, the attacker employs the well-known tool Nmap\footnote{\url{https://nmap.org/}} to carry out DNS and port scans in the DMZ network where the VPN server is located. This allows the attacker to discover the CIDR of the intranet network and thus extend their scans to the hosts located in the intranet zone. Eventually, a web service scan shows a WordPress instance running on the intranet server, which leads to the attacker selecting this server as a possible target for intrusion. The attacker thus launches a brute force directory scan using the tool dirb\footnote{\url{https://tools.kali.org/web-applications/dirb}} in order to find potentially interesting files. Since this scan shows up no results that allow the attacker to progress any further, they carry out a WordPress security scan using the tool WPScan\footnote{\url{https://wpscan.com/wordpress-security-scanner}} in order to discover vulnerable versions or misconfigurations of plugins or themes installed on the server. Other than the directory scan, this security scan shows that a vulnerable version of the plugin wpDiscuz is present on the server. At this point, the attacker stops scanning and instead focuses on exploiting the vulnerability, which marks the end of the reconnaissance phase.

\begin{sidewaystable*}[]
	\centering
	\footnotesize
	\caption{Overview of the attack scenario}
	\begin{tabular}[t]{lllll} \hline
		\textbf{Kill chain phases} & \textbf{Attack steps} & \textbf{Tools} & \textbf{MITRE ATT\&CK Tactics and Techniques} & \textbf{Data sources} \\ \hline \hline
		Reconnaissance & \makecell[lt]{ Traceroute \\ Network scan \\ DNS scan \\ Service scan } & Nmap & \makecell[lt]{ Reconnaissance \\ - Active Scanning \\ - Gather Victim Network Information  } & \makecell[lt]{ DNS logs \\ Network traffic } \\ \hline
		Reconnaissance & \makecell[lt]{ WordPress scan \\ Directory scan } & \makecell[lt]{ WPScan \\ Dirb } & \makecell[lt]{ Reconnaissance \\ - Active Scanning \\ - Gather Victim Host Information  } & \makecell[lt]{ Access logs \\ Error logs \\ Network traffic } \\ \hline
		\makecell[lt]{ Initial Intrusion \\ Establish a Backdoor } & \makecell[lt]{ Webshell upload \\ Webshell command execution } & Shell & \makecell[lt]{ Execution \\ - Exploitation for Client Execution \\ Persistence \\ - Server Software Component \\ Discovery } & \makecell[lt]{ Access logs } \\ \hline
		\makecell[lt]{ Obtain User Credentials } & \makecell[lt]{ Wordpress database dump } & Shell & \makecell[lt]{ Credential Access \\ - OS Credential Dumping } & \makecell[lt]{ Access logs } \\ \hline
		\makecell[lt]{ Obtain User Credentials \\ Install Various Utilities } & \makecell[lt]{ Password cracking } & \makecell[lt]{ John the \\ Ripper } & \makecell[lt]{ Credential Access \\ - Brute Force: Password Cracking } & \makecell[lt]{ Monitoring logs } \\ \hline
		\makecell[lt]{ Privilege Escalation } & \makecell[lt]{ Login as system user } & Shell & \makecell[lt]{ Privilege Escalation \\ - Valid Accounts } & \makecell[lt]{ Auth logs \\ Audit logs } \\ \hline
		\makecell[lt]{ Lateral Movement } & \makecell[lt]{ Reverse shell setup \\ Root command execution } & Shell & \makecell[lt]{ Execution \\ - Command and Scripting Interpreter } & \makecell[lt]{ Auth logs \\ Audit logs } \\ \hline
		\makecell[lt]{ Data Exfiltration } & \makecell[lt]{ Exfiltration over DNS } & DNSteal & \makecell[lt]{ Exfiltration \\ - Exfiltration Over Alternative Protocol  } & \makecell[lt]{ DNS logs \\ Audit logs } \\ \hline
	\end{tabular}
	\label{tab:mapping}
\end{sidewaystable*}

\begin{table}[]
	\centering
	\footnotesize
	\caption{Variations of the attack scenario}
	\begin{tabular}[t]{lll} \hline
		\textbf{Attack} & \textbf{Parameter} & \textbf{Range} \\ \hline \hline
		General & Start times & 00:00 - 24:00 \\
		& Attacker name & Random name \\ \hline
		Network scans & Ports & 100-2000 top ports \\
		& Hosts & Random selection of servers \\ \hline
		Wordpress scan & Scan mode & {[ passive, mixed ]} \\
		& Enumeration & \makecell[lt]{Random selection of plugins, \\ themes, configs., database \\ exports, users, and media} \\ \hline
		Directory scan & Recursive & {[ yes, no ]} \\
		& Case-sensitive & {[ yes, no ]} \\ \hline
		Webshell & Shell name & Random string \\
		& Commands & Random commands \\ \hline
		\makecell[lt]{ Password hash } & Mode & {[ online, offline ]} \\ 
		cracking & Duration & 30-90 minutes \\ \hline
		Reverse shell & Port & 1100-65000 \\
		& Commands & Random commands \\ \hline
		Exfiltration & DNS domain & Random string \\
		& Forced IP & {[ yes, no ]} \\
		& Compression & {[ yes, no ]} \\ 
		& Verbosity & {[ yes, no ]} \\ 
		& Block size & 32-63 \\
		& Sub domains & integer of (200 / block size) \\ \hline
	\end{tabular}
	\label{tab:attacks}
\end{table}

By exploiting the vulnerable plugin, the attacker is able to perform unrestricted file uploads (CVE-2020-24186). This allows the attacker to upload a PHP webshell as a backdoor that in turn allows them to execute arbitrary commands with the privileges of the \textit{www-data} user of the web server. The attacker proceeds to execute several commands to gather information about the host, e.g., reading out processes, command histories, OS information, connections, or file names. Eventually the attacker finds the password to the user database in the WordPress configuration file and is thus able to access all user names and their hashed passwords.

The attacker then attempts to crack one of the hashed passwords using a list of common passwords. For this, our attacker state machine branches into two paths. In one path, we assume that the attacker transfers the password hashes to their own system and manages to crack one of the passwords there. Since this activity takes place outside of the monitored network, no logs are created and thus detection is not possible. Accordingly, we simulate this case by simply pausing the state machine for a specific amount of time. The other path simulates that cracking takes place at the compromised server. For this, the attacker installs the tool John the Ripper\footnote{\url{https://www.openwall.com/john/}} and uses a common password list for cracking. Due to the fact that the purpose of our datasets is to provide detectable traces of anomalous behavior, we opt for the latter case when running our simulations. Note that as part of our attack scenario, we assume that the password of at least one system user is always present in the password list and thus successfully cracked after a certain amount of time. Subsequently after obtaining the password, the attacker uploads a fully interactive reverse shell and misuses the compromised user account to escalate their privileges to root level. The attacker then executes several commands of which some require root privileges, such as reading out the shadow file.

As a final step of the attack kill chain, the attacker runs the DNSteal\footnote{\url{https://github.com/m57/dnsteal}} tool that exfiltrates sensitive data from the file share located in the intranet zone. Thereby, the tool starts a process that converts files from certain directories into base64 to conform to the requirements of DNS queries, splits them into chunks, and sends them as DNS requests through the firewall to a specific attacker-controlled domain in the global DNS. Eventually the data is transferred from the malicious domain to the attacker's host, where it is decoded and stored. While we could have modeled the attack chain in a way so that the attacker would set up this exfiltration tool once they gained system privileges, we decided to separate this step from the remaining attack vectors and instead start the exfiltration tool already at the beginning of the simulation. The reason for this is that we decided to design the exfiltration attack as a challenge for anomaly-based IDSs that usually rely on an training phase that is free of attacks. By running the tool from the beginning of the simulation, we purposefully poison the training phase so that the malicious DNS communication is learned as part of the normal system behavior. However, the attack may still be detected by anomaly-based IDSs, since the exfiltration stops after a few days when all files are extracted. This is especially challenging, since it is usually more difficult for an IDS to recognize that a service suddenly stopped compared to the detection of a newly started service.

Table \ref{tab:mapping} summarizes the attack scenario. The first column maps each of the attack steps stated in the second column to phases of the cyber kill chain \cite{fireeyeapt}. As stated before, the \textit{Data Exfiltration} step does not chronologically follow the other attack steps. The third column lists related tactics and techniques from the well-known MITRE ATT\&CK matrix version 10 \cite{mitreattack} for each attack step. The matrix classifies and describes a wide range of common attack techniques and also provides information on detection. As visible in the table, our multi-step attack involves a diverse set of attack techniques that are part of several tactics. Finally, the last column states the most relevant log files that contain attack traces for each attack step. Since many different log files are affected, it is necessary to configure IDSs to monitor several hosts of the network in order to obtain a full picture of the multi-step intrusion.

Similar to the infrastructure and user behavior, we vary the attack parameters as part of the transformation from TIM to TSM. Table \ref{tab:attacks} provides an overview of the main variations used to generate the dataset. Note that while the time of day at which attack execution is initiated is varied, we manually set the day for each simulation run in advance. The reason for this is to avoid that the attack is launched too early and thus the dataset does not provide a sufficiently long training phase of at least 3 days. To select and implement variations of parameters of utilized attack tools, we looked up allowed values and ranges for each parameter in the respective documentations. Since tools such as WPScan and DNSteal have multiple parameters that support ranges of allowed values, many possible combinations of values exist and thus the attack traces resulting in the logs are highly different. To realize random command executions, we assembled a list of common commands and randomly sampled them. We also injected the user password to be cracked in specific positions of the password file used by John the Ripper so that the duration to complete cracking varies in each run.

\section{Analysis of Log Datasets} \label{analysis}

The previous section outlined our methodology and scenario for generating testbeds using a model-driven approach. Following this methodology, we generated eight testbeds and collected log data from them. This section provides some insights into these datasets by analyzing and comparing the logs.

\subsection{Testbed Infrastructures}

In course of around four weeks we instantiated a total of eight testbeds that we used to collect log datasets. The durations of the simulations for each dataset are between 4-6 days, where the exfiltration attack that is already running in the beginning of the simulation usually stops after 1-3 days and the multi-step server takeover attack usually takes place one of the last two days. 

Table \ref{tab:overview} provides an overview of the technical infrastructure used to generate each of the datasets. Note that we refer to each dataset by the randomly selected name of the overall testbed network that contains all zones. As visible in the table, the randomly selected numbers of mail servers and user host machines present in the testbeds correspond to the parameter variations stated in Sect. \ref{usecase}. We point out that the size of the datasets mostly depend on the number of active users and the length of the simulation.

\begin{sidewaystable*}[]
	\footnotesize
	\centering
	\caption{Technical infrastructure of testbeds}
	\begin{tabular}[t]{llccccccc}
		\hline
		\textbf{Dataset} & \textbf{Network} & \makecell[cb]{ \textbf{Mail} \\ \textbf{servers} } & \makecell[cb]{ \textbf{Internal} \\ \textbf{employees} } & \makecell[cb]{ \textbf{Remote} \\ \textbf{employees} } & \makecell[cb]{ \textbf{External} \\ \textbf{users} } & \textbf{Start} & \textbf{End} & \textbf{Duration} \\ \hline \hline
		fox & fox.org & 4 & 5 & 4 & 7 & 2022-01-15 00:00 & 2022-01-20 00:00 & 5 days \\ \hline
		harrison & harrison.com & 2 & 3 & 6 & 6 & 2022-02-04 00:00 & 2022-02-09 00:00 & 5 days \\ \hline
		russellmitchell & russellmitchell.com & 2 & 4 & 3 & 3 & 2022-01-21 00:00 & 2022-01-25 00:00 & 4 days \\ \hline
		santos & santos.com & 2 & 9 & 3 & 6 & 2022-01-14 00:00 & 2022-01-18 00:00 & 4 days \\ \hline
		shaw & shaw.info & 3 & 5 & 5 & 3 & 2022-01-25 00:00 & 2022-01-31 00:00 & 6 days \\ \hline
		wardbeck & wardbeck.info & 3 & 6 & 7 & 4 & 2022-01-19 00:00 & 2022-01-24 00:00 & 5 days \\ \hline
		wheeler & wheeler.biz & 4 & 8 & 6 & 8 & 2022-01-26 00:00 & 2022-01-31 00:00 & 5 days \\ \hline
		wilson & wilson.com & 2 & 7 & 8 & 9 & 2022-02-03 00:00 & 2022-02-09 00:00 & 6 days \\ \hline
	\end{tabular}
	\label{tab:overview}
\end{sidewaystable*}

Table \ref{tab:files} shows which log files are collected from which hosts, where \checkmark indicates that the respective log file is collected from the host, \textcircled{$\checkmark$} indicates that the respective log file is collected and also labels exist for that file, and no symbol indicates that the respective files are not collected or not present on the hosts. The table also shows that we collect network traffic as well as system logs from diverse sources, for example, access logs, low-level logs of the operating system (audit logs), application logs (Horde and VPN logs), monitoring logs, custom logs for state machine executions, etc. Note that files not marked as labeled do not necessarily lack a ground truth, since several files are not affected by any of the attacks and thus all occurring events correspond to normal behavior. We therefore only mark files as labeled in case that attack traces are known to occur in these files and labeling rules for the respective attack manifestations exist.

As visible in the table, we mainly focused on log files from the intranet server when developing our labeling rules. The reason for this is that the majority of attack steps are launched against that server and the diversity of these attack vectors cause that several different files are affected. In Sect. \ref{labels} we provide a more detailed overview of assigned labels. 

\newcommand*\rot{\rotatebox[origin=l]{90}}
\begin{sidewaystable*}[]
	\footnotesize
	\centering
	\caption{Log files collected from hosts}
	\begin{tabular}[t]{lcccccccccccccccccccccc}
		%\hline
		Host & \rot{Statemachine logs} & \rot{Network traffic} & \rot{Apache access logs} & \rot{Apache error logs} & \rot{Auth logs} & \rot{Journal logs} & \rot{DNS logs} & \rot{VPN logs} & \rot{Syslog} & \rot{Audit logs} & \rot{Suricata event logs} & \rot{Suricata fast logs} & \rot{Suricata stats} & \rot{Kernel logs} & \rot{Exim logs} & \rot{Horde access logs} & \rot{Horde error logs} & \rot{Mail (info) logs} & \rot{Mail warning logs} & \rot{Messages} & \rot{User logs} & \rot{Monitoring logs} \\ \hline \hline
		attacker host & \checkmark & \checkmark &  &  &  &  &  &  &  &  &  &  &  &  &  &  &  &  &  &  & & \\ \hline
		employee host & \checkmark &  &  &  &  &  &  &  &  &  &  &  &  &  &  &  &  &  &  &  & & \\ \hline
		intranet server &  & \checkmark & \textcircled{$\checkmark$} & \textcircled{$\checkmark$} & \textcircled{$\checkmark$} & \checkmark &  & & \checkmark & \textcircled{$\checkmark$} & \checkmark & \checkmark & \checkmark &  &  &  &  &  &  &  &  &  \textcircled{$\checkmark$} \\ \hline
		file share &  & \checkmark & \checkmark & \checkmark & \checkmark & \checkmark & & & \checkmark & \textcircled{$\checkmark$} & \checkmark & \checkmark & \checkmark &  &  &  &  &  &  &  &  &  \checkmark \\ \hline
		int. mail server &  & \checkmark & \checkmark & \checkmark & \checkmark &  &  & & \checkmark & \checkmark & \checkmark & \checkmark & \checkmark &  & \checkmark & \checkmark & \checkmark & \checkmark & \checkmark & \checkmark & \checkmark &  \\ \hline
		ext. mail server &  &  &  &  & \checkmark &  &  & & \checkmark &  &  &  &  &  & \checkmark & \checkmark & \checkmark & \checkmark & \checkmark & \checkmark & \checkmark &  \\ \hline
		firewall &  & \checkmark &  &  & \checkmark & \checkmark & \textcircled{$\checkmark$} & & \checkmark & \checkmark & \checkmark & \checkmark & \checkmark & \checkmark &  &  &  &  &  &  &  &  \\ \hline
		DNS server &  &  &  &  & \checkmark & \checkmark & \checkmark &  & \checkmark &  &  &  &  &  &  &  &  &  &  &  &  & \\ \hline
		VPN server &  & \checkmark &  &  & \checkmark & \checkmark &  & \textcircled{$\checkmark$} & \checkmark & \checkmark & \checkmark & \checkmark & \checkmark &  &  &  &  &  &  &  &  & \\ \hline
		web server &  & \checkmark & & & \checkmark & \checkmark &  &  & \checkmark & \checkmark & \checkmark & \checkmark & \checkmark &  &  &  &  &  &  &  &  & \\ \hline
		cloud share &  & \checkmark &  &  & \checkmark & \checkmark & & & \checkmark & \checkmark & \checkmark & \checkmark & \checkmark &  &  &  &  &   &  &  &  & \\ \hline
	\end{tabular}
	\label{tab:files}
\end{sidewaystable*}

\subsection{Normal Behavior} \label{normal}

As pointed out in Sect. \ref{requirements}, it is essential for synthetic log data generation to simulate normal user behavior that corresponds to real humans interacting with the system in terms of click frequency as well as complexity and diversity of actions. However, we noticed in our literature review (cf. Sect. \ref{literature}) that comparisons of presented datasets with real user behavior are rarely carried out. We therefore validate our log datasets by carrying out a comparison with real-world log data generated by humans performing tasks in a similar network environment. The real log data was collected during a cyber security exercise\footnote{Austrian Press Agency, \url{https://www.ots.at/presseaussendung/OTS_20210922_OTS0036}} that took place in September of 2021. As part of the exercise, eight teams consisting of four people respectively were tasked to investigate traces of existing malware that infected their networks, monitor their systems for incoming cyber attacks, and respond to incidents by contacting authorities. As part of this one-day exercise, several attacks were scheduled for automatic execution at specific points in time, keeping the participants busy at all times. During the exercise, the teams worked isolated from each other and could not access the technical infrastructure of other teams. 

\begin{figure*}[]
	\centering
	\begin{subfigure}[t]{0.5\textwidth}
		% 12x12
		\centering
		\includegraphics[width=1\columnwidth]{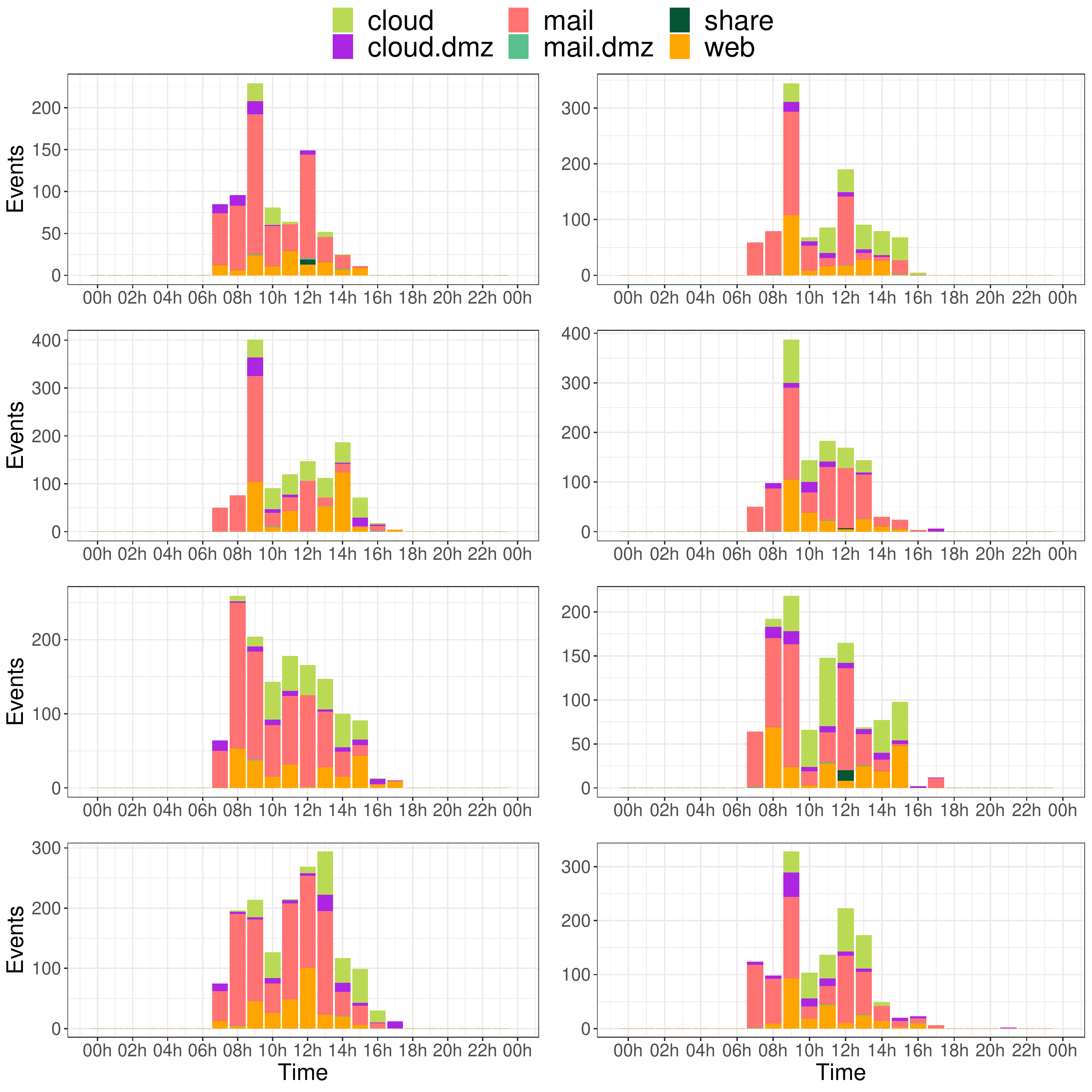}
		\caption{Access frequencies of real users.}
		\label{fig:aaa}
	\end{subfigure}%
	~ 
	\begin{subfigure}[t]{0.5\textwidth}
		% 12x12
		\centering
		\includegraphics[width=1\columnwidth]{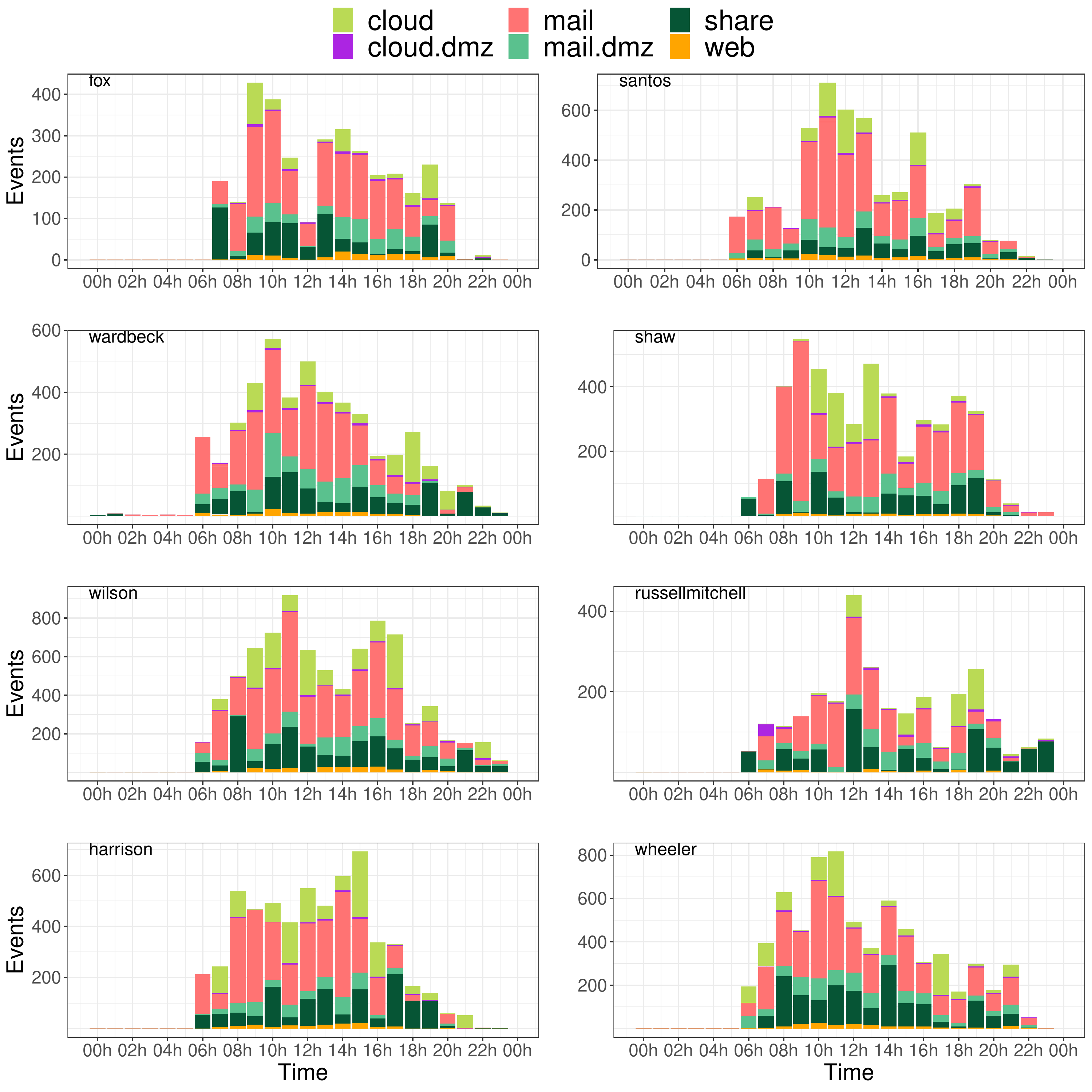}
		\caption{Access frequencies of simulated users.}
		\label{fig:bbb}
	\end{subfigure}
	\caption{Event counts in DNS logs for different services.}
	\label{fig:comp}
\end{figure*}

To set up the system environment for each team, most of the provisioning scripts were reused as TIMs for setting up the testbed as outlined in Sect. \ref{usecase}. This allows us to compare the contents of the log files generated in the environments utilized by real humans and those of our dataset. We select the DNS logs as a base for comparison, since they contain queries on a level of abstraction that allows us to determine whether users accessed the cloud server, mail server, file share, etc. Figure \ref{fig:comp} visualizes the events produced of the real users (left) and simulated users (right). Note that we only use logs from the first day of each dataset since there is also just one day of logs from real users available.

The plots show that there are some discrepancies between real and simulated users, however, these are mostly linked to some conscious design decisions. First, it is apparent that logs generated by simulated users are more spread out across the day with logs occurring between 5:00-22:00, while real users only produced logs between 7:00-17:00. This is clearly caused by the fact that the cyber security exercise had a clear start and end time and participants were not freely able to carry out their tasks at any time they desire. Accordingly, we argue that the user behavior in our datasets that simulates employees rather than participants of an exercise adequately represents the active times of employees with flexible working hours. Similarly, real logs show that users hardly ever accessed the file share, which is mostly due to the fact that none of their tasks were linked to sharing files with each other. Overall, the relative frequencies of accesses per service from real users largely resemble those of simulated users, with mail servers being the most actively accessed services. Considering the absolute event frequencies, the simulation appears to correctly depict access frequencies of real users in terms of average accesses per person and hour as well as fluctuations thereof across the day. In particular, we computed that real users generate 306.2 DNS events per day across all services on average with a standard deviation of 62.2 and simulated users generate 307.2 DNS events per day across all services on average with a standard deviation of 56.7.

\subsection{Attacks} \label{attacks}

Manifestations of attack executions in log data and labels thereof are crucial for log datasets. As discussed in Sect. \ref{att_sce}, we designed our attack scenario to involve a wide variety of attack types that affect several different files. In the following, we exemplarily show how some of these attack steps manifest themselves in the generated datasets.

\begin{figure}
	% 12x6
	\centering
	\includegraphics[width=.8\columnwidth]{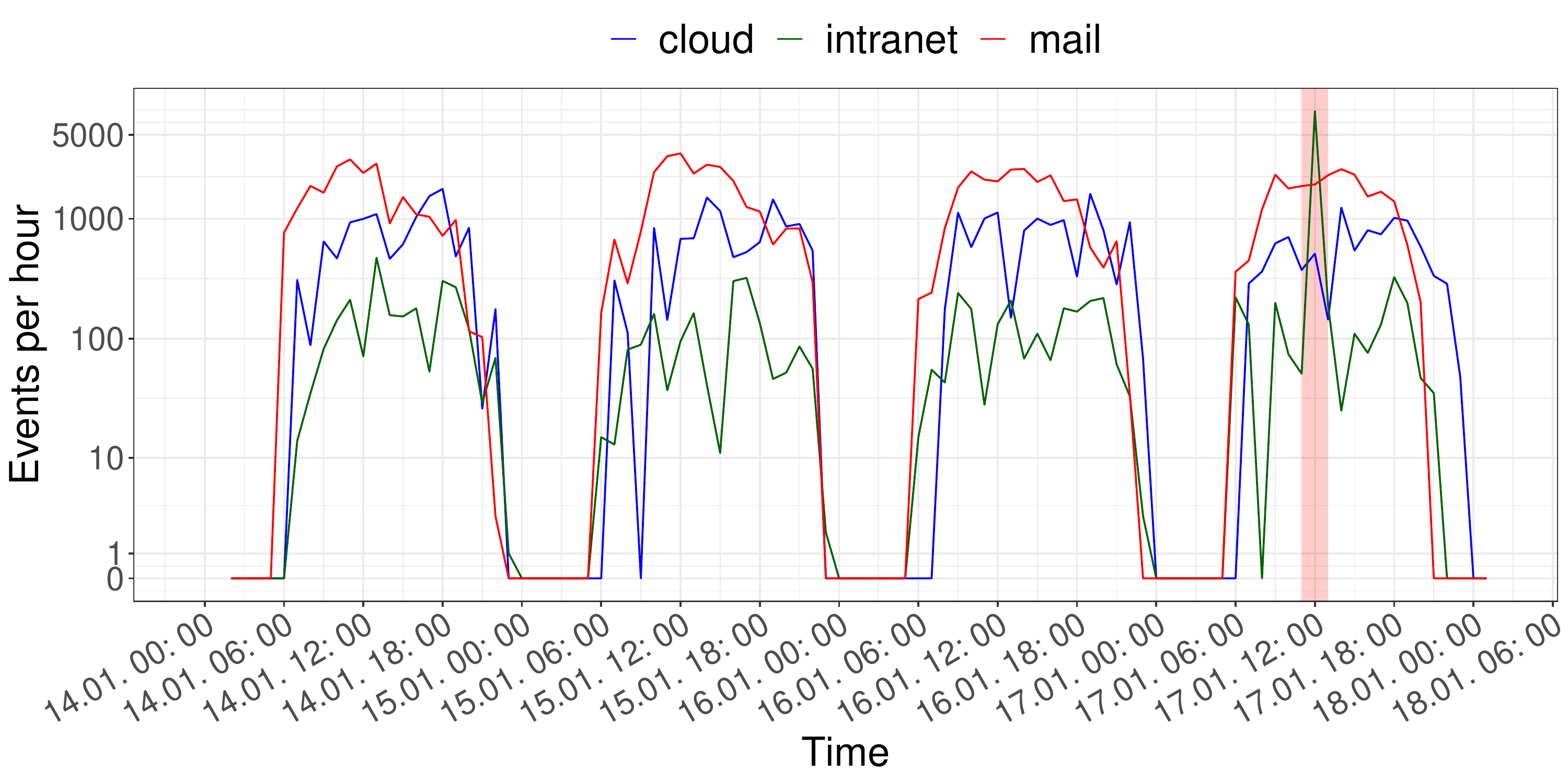}
	\caption{Apache access logs with attack consequences of scans.}
	\label{fig:apache}
\end{figure}

One of the most recognizable attack steps is the directory scan that is carried out as part of the reconnaissance phase. This attack makes several thousands of requests in a short amount of time to the targeted web server, of which all are recorded in the Apache access logs. Since this log file usually contains events that relate to users requesting resources by clicking around on web pages, the scan causes a drastic increase of the average load during normal system operation. Figure \ref{fig:apache} shows the number of events in the Apache access logs per hour on the cloud, intranet, and mail servers of the \textit{santos} dataset. As visible in the plot, the accesses on the intranet server during the directory scan (the relevant time interval is shaded red) increase from several hundred to more than 5000.

\begin{figure}
	% 12x6
	\begin{subfigure}{1\columnwidth}
		\centering
		\includegraphics[width=.69\columnwidth]{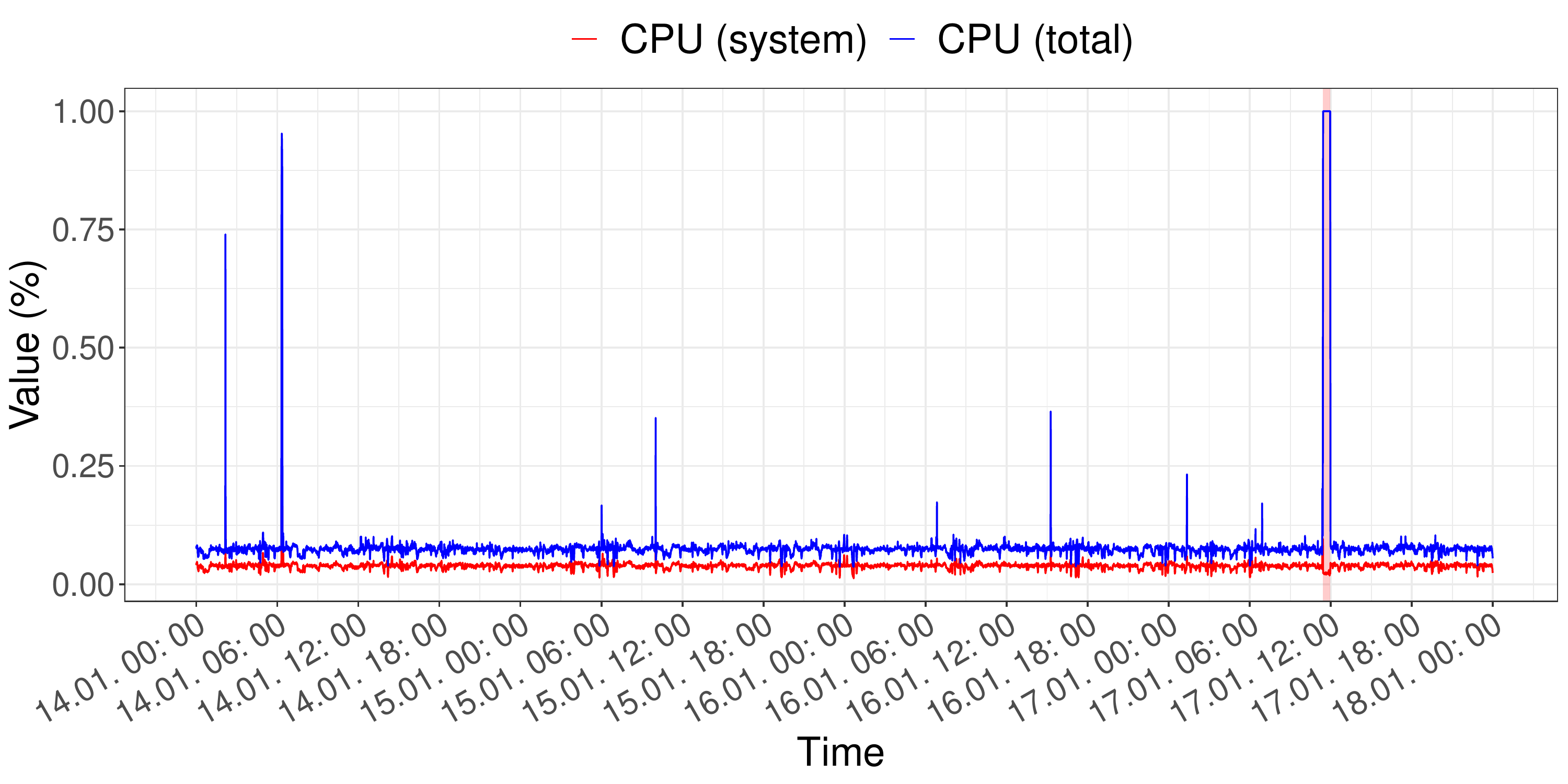}
	\end{subfigure}
	\begin{subfigure}{1\columnwidth}
		\centering
		\includegraphics[width=.69\columnwidth]{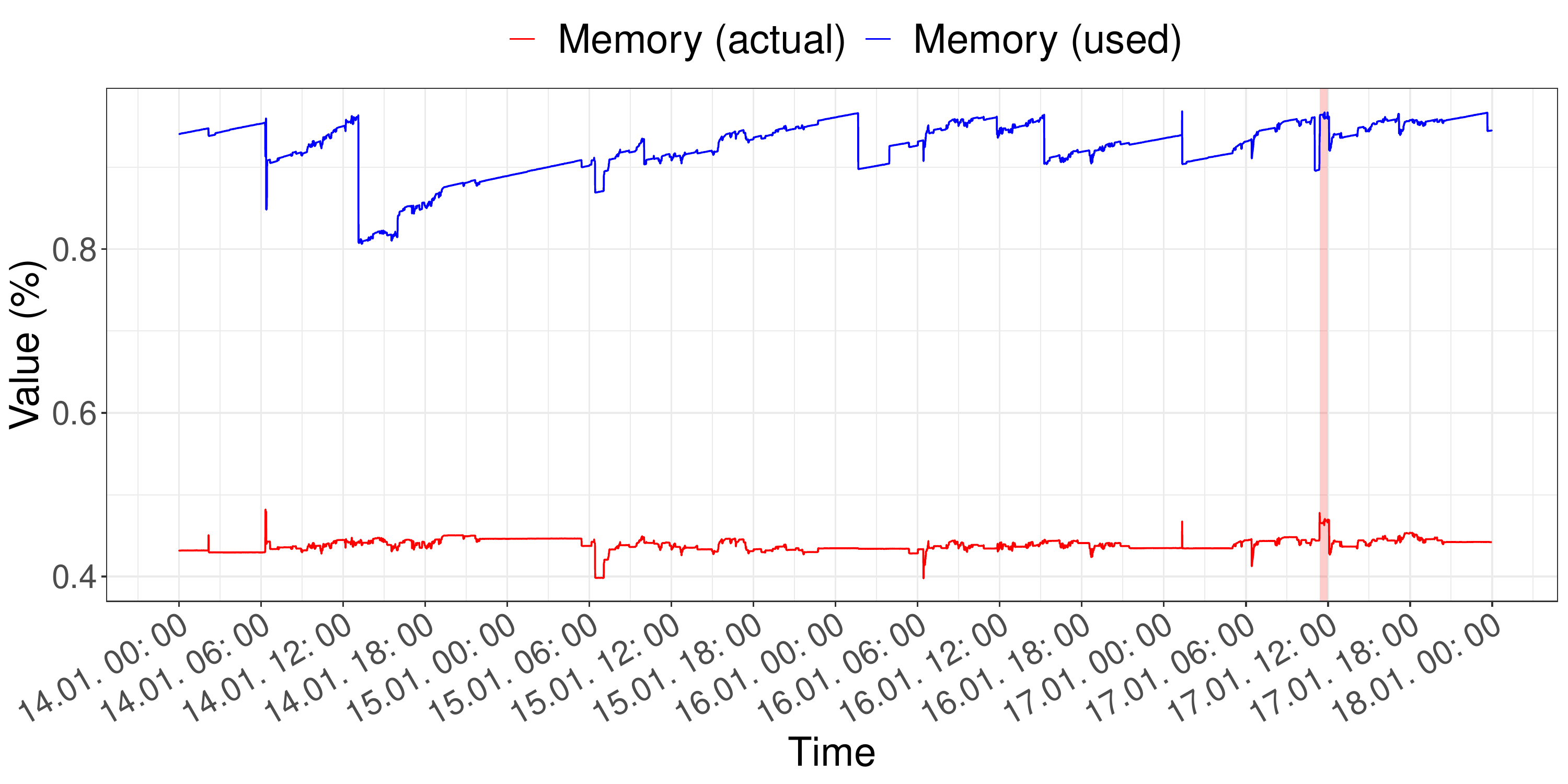}
	\end{subfigure}
	\caption{Monitoring logs of CPU (top) and memory (bottom) showing attack consequences of password cracking.}
	\label{fig:monitoring}
\end{figure}

Monitoring logs contain numeric values of system measurements that are an interesting input for anomaly detection \cite{khorshed2011monitoring}. This includes measurements on the utilization of CPU, memory, disk, file system, network communication, processes, etc. For our datasets, we collect such monitoring logs from the file share and intranet server that are both located in the intranet zone and are thus reasonable targets for monitoring in real-world scenarios. Figure \ref{fig:monitoring} shows several metrics derived from CPU and memory utilization that are collected from the \textit{santos} dataset. As visible in the top plot, both system and total CPU are significantly increased as a consequence of the password cracking attack step (the relevant time interval is shaded red). The memory metrics do not show such a strong indication of an ongoing attack, even though a large file containing passwords is loaded into memory during cracking. Nonetheless, these and other metrics or combinations thereof could also contribute to the detection of certain attack steps.

Variations of the system environment, normal behavior simulation, and attack parameters, cause that aforementioned attack consequences differ across datasets. For example, peaks in event frequencies have different magnitudes relative to the baseline of event occurrences that is considered normal for that dataset and the time intervals where system metrics are affected change in length. In addition, event sequences that are generated as a consequence of commands executed by the attacker have different form or parameters. Consider the log events shown in Fig. \ref{fig:auth} as an example. In the \textit{fox} dataset (top), seven events are generated when the attacker logs into the compromised user account \textit{phopkins}. The same attack step appears different in the \textit{harrison} dataset, as both the affected user changes to \textit{jward}, terminal \textit{/dev/pts/0} rather than \textit{/dev/pts/1} is used, and different commands are executed. We argue that these variations are useful to achieve higher robustness of results when evaluating IDSs, since detection accuracy should be similar across all datasets even though the events to be detected vary. In Sect. \ref{application}, we discuss the benefits of our datasets in more detail.

\begin{figure*}
\centering
\scriptsize
\begin{Verbatim}
Jan 18 13:14:31 intranet-server su[28816]: Successful su for phopkins by www-data
Jan 18 13:14:31 intranet-server su[28816]: + /dev/pts/1 www-data:phopkins
Jan 18 13:14:31 intranet-server su[28816]: pam_unix(su:session): session opened for user
  phopkins by (uid=33)
Jan 18 13:14:31 intranet-server systemd-logind[1011]: New session c1 of user phopkins.
Jan 18 13:14:31 intranet-server systemd: pam_unix(systemd-user:session): session opened for
  user phopkins by (uid=0)
Jan 18 13:14:41 intranet-server sudo: phopkins : TTY=pts/1 ; USER=root ; COMMAND=list
Jan 18 13:14:43 intranet-server sudo: phopkins : TTY=pts/1 ; USER=root ; COMMAND=
  /bin/ls -laR /root/
		
Feb  8 08:36:38 intranet-server su[28321]: Successful su for jward by www-data
Feb  8 08:36:38 intranet-server su[28321]: + /dev/pts/0 www-data:jward
Feb  8 08:36:38 intranet-server su[28321]: pam_unix(su:session): session opened for user
  jward by (uid=33)
Feb  8 08:36:38 intranet-server systemd-logind[935]: New session c1 of user jward.
Feb  8 08:36:38 intranet-server systemd: pam_unix(systemd-user:session): session opened for
  user jward by (uid=0)
Feb  8 08:36:54 intranet-server sudo:    jward : TTY=pts/0 ; USER=root ; COMMAND=list
Feb  8 08:36:57 intranet-server sudo:    jward : TTY=pts/0 ; USER=root ; COMMAND=
  /bin/cat /etc/shadow
\end{Verbatim}
\caption{Different log events caused by the attacker escalating to system privileges in the \textit{fox} (top) and \textit{harrison} (bottom) datasets.}
\label{fig:auth}
\end{figure*}

\subsection{Labels} \label{labels}

As explained in Sect. \ref{testbed_gen}, our labeling procedure does not just make use of attack time windows to mark events as malicious based on their timestamps, but instead involves query rules that enable labeling based on event attributes. We created such rules for eight files as outlined in Table \ref{tab:files} and assign distinct labels to malicious events based on their attack step. Note that we specifically selected files and attack steps which involve distinct manifestations of attack consequences after manually checking all files, however, we also point out that there are traces of attack steps in other files that are not labeled in AIT-LDSv2.0. Due to the fact that our collection of log datasets is maintainable and the labeling procedure is repeatable, it is possible to add labeling rules for these files in future versions of the dataset. 

\begin{figure}
	% 6x4
	\centering
	\includegraphics[width=.7\columnwidth]{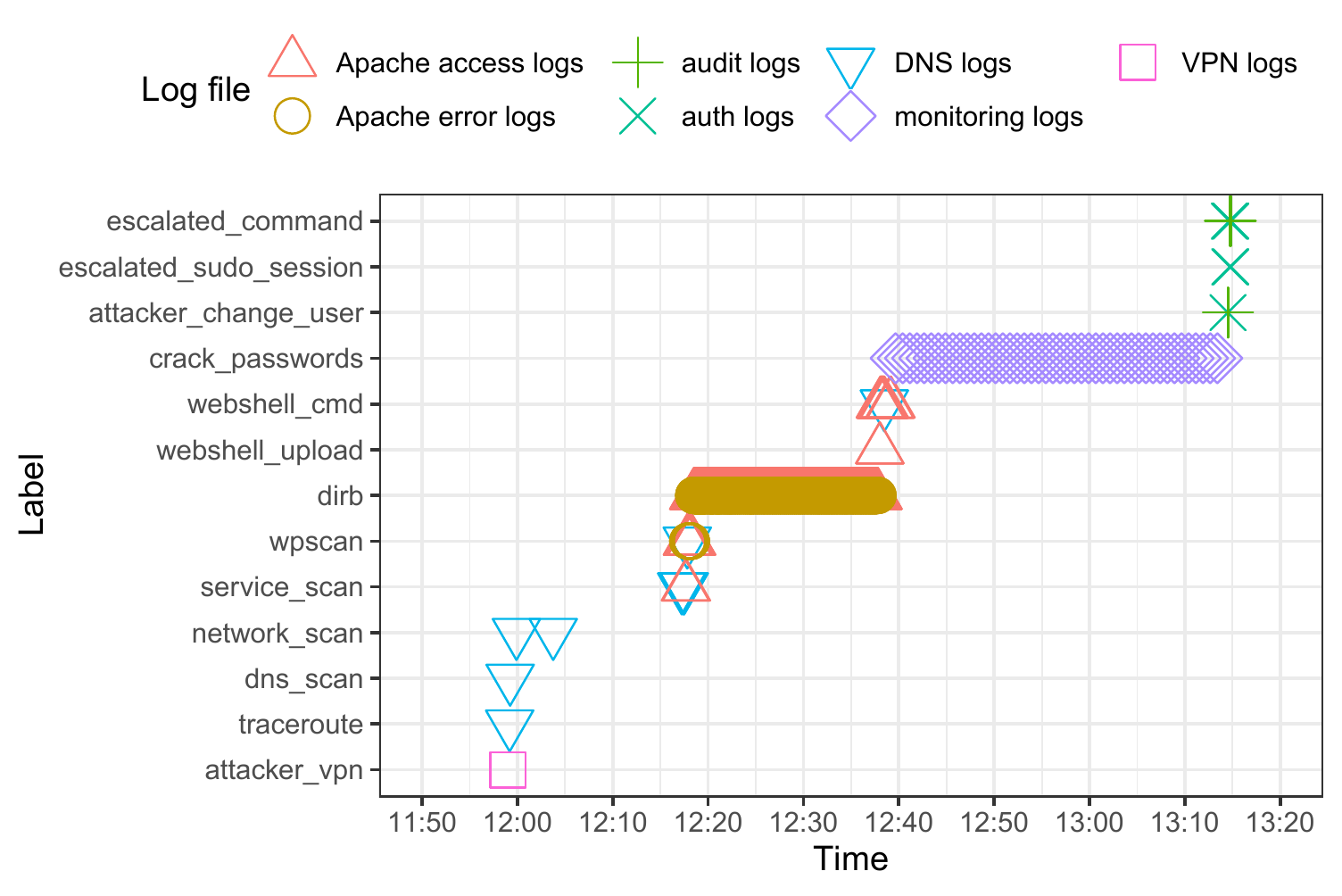}
	\caption{Occurrences of events labeled as part of the multi-step attack.}
	\label{fig:labels}
\end{figure}

We exemplarily show an overview of labeled events related to the multi-step attack of the \textit{santos} dataset in Fig. \ref{fig:labels}. The figure visualizes the chronological occurrence of labeled events, where the distinct labels are depicted on the vertical axis and affected files are marked with different symbols. As visible in the plot, some attack steps cause singular events or short sequences (e.g., uploading the webshell), while others affect groups of events that span over a longer duration (e.g., password cracking). Note that we assign multiple labels to the same events for clarification. For example, we introduce a label \textit{foothold} that subsumes all attack steps involved in the initial intrusion, including the VPN connection, scans, and webshell upload. This implies that our labels follow a hierarchical order, which makes it easy to select specific types of events for evaluation and furthermore allows to compute detection accuracies separately for different attack steps \cite{landauer2020kyoushi}.

\section{Discussion} \label{discussion}

In this section we discuss whether our generated datasets fulfill the requirements for IDS evaluation. In addition, we explain possible application scenarios for our datasets in detail and outline their limitations.

\subsection{Fulfillment of Requirements} \label{fulfillment}

We stated requirements for log dataset generation that we used as a basis for our methodology in Sect. \ref{requirements}. Based on the generated datasets and the results of our analysis provided in Sect. \ref{analysis} we check whether all requirements are fulfilled. Requirement (1) is fulfilled as our datasets address enterprise IT, which is a wide-spread and relevant use-case for intrusion detection. We followed common guidelines for network design and selected open-source components that are popular choices in such infrastructures \cite{macia2018ugr}. Requirement (2) addresses simulation of normal system behavior. We argue that our state machines and randomized user role assignments that are used for simulating employees as outlined in Sect. \ref{usersim} are sufficiently extensive to generate complex patterns. Moreover, we show in Sect. \ref{normal} that page visit frequencies of our simulated employees largely resemble those of real users. Similarly, our selected attack scenarios involve diverse steps and recent exploits to fulfill requirement (3). We collect both system log data as demanded by requirement (4) as well as network traffic as demanded by requirement (5). Our user simulations follow daily activity cycles as visible in our analysis results presented in Sect. \ref{attacks}. Since multiple days of such normal behavior is recorded, we consider requirement (6) that addresses periodic patterns as fulfilled. Requirement (7) is fulfilled as we generate a ground truth for events using our labeling framework as described in Sect. \ref{testbed_gen}. Beside a description of the overall scenario available in this paper and the dataset repository, all scripts and configurations of our testbeds are published together with the log data and thus also requirement (8) on the availability of documentation is fulfilled. Requirements (9) and (10) are fulfilled, because we generate multiple datasets that contain repeated executions of the same attack steps with variations. Finally, requirement (11) is fulfilled as we publish all scripts for deploying and running the simulation as open-source code.

\subsection{Application Scenarios for AIT-LDSv2.0} \label{application}

Due to the characteristics of our dataset, we foresee several different application scenarios. In the following, we discuss (federated) intrusion detection, alert aggregation, and user profiling as interesting research areas that benefit from our data.

\subsubsection{Evaluation of Intrusion Detection Systems}

Foremost, the purpose of our collection of datasets is to enable evaluation of host- and network-based IDSs. We injected attacks that employ diverse techniques so that their consequences manifesting in log files challenge a wide range of detection mechanisms \cite{skopik2021smart}. For example, we anticipate the following non-exhaustive list of detection techniques to be applied on our dataset. 

\begin{itemize}
	\item \textbf{New log artifacts}. As part of many attack steps, new log events such as the sample logs from Fig. \ref{fig:auth} appear in some log files. Alternatively, normal event types may appear with different parameters or combinations of parameter values. Despite the fact that this detection technique is relatively simple, it is highly powerful, because its low runtime requirements can be applied to most events or categorical values. 
	\item \textbf{Structure of parameter values}. The DNSteal attack makes use of a randomly generated domain names for data exfiltration, which could be useful to evaluate detectors for domain-generation algorithms \cite{woodbridge2016predicting}. The same applies for Apache access logs, where commands sent to the webshell appear in URLs. 
	\item \textbf{Sequence mining}. Log events usually occur in specific sequences that represent inherent program flows of monitored services. Workflow mining extracts these patterns and allows to detect unusual sequences as anomalies \cite{du2017deeplog, he2016experience}. Consequences of exploits and other malicious attacker behavior often manifest in such sequences, for example, audit events generated when the attacker executes commands via the remote shell. 
	\item \textbf{Event frequencies}. As pointed out in Sect. \ref{attacks}, attacks such as scans are recognizable by high amounts of log occurrences in short time intervals. Anomaly detection techniques therefore create event count matrices and detect time windows with unusual high or low event frequencies with the aid of various machine learning methods, including time-series analysis \cite{landauer2018dynamic} and principal component analysis \cite{he2016experience}.
	\item \textbf{Missing events}. We deliberately designed our attack scenario to include a data exfiltration attack that is already ongoing at the beginning of the simulation and stops after some days. We expect that detectors based on machine learning add these malicious events to their model of normal behavior that is generated during the training phase, and thus poison their models. Accordingly, detectors need to raise anomalies for the stopping of event occurrences, which we consider a more challenging detection scenario than recognizing the start of the exfiltration process.
	\item \textbf{Statistical tests}. System performance metrics and numeric features of network traffic are suitable for statistical analyses such as testing for certain distributions. Alternatively, hypothesis testing is also applicable for detecting changes of correlating behavior of categorical variables in log data \cite{landauer2021iterative}. 
\end{itemize}

We argue that our data has a large benefit over most existing datasets for IDS evaluation, as it contains data from multiple separate testbeds targeted by the same attack scenario. Due to the variations in the log traces caused by changes of the system environment, simulated normal behavior, and attack parameters, we expect that detection accuracies vary when applying the same detectors on different datasets. However, by averaging the detection metrics achieved on all datasets, the aggregated results have a higher robustness as they are more representative for a general case and not fine-tuned to only a single execution. In addition, simulating many similar infrastructures allows to evaluate approaches that leverage federated learning for intrusion detection \cite{preuveneers2018chained}.

Moreover, the ground truth tables of our datasets are not just binary labels that determine whether an event is part of an attack or not, but instead precisely state the type of attack. This means that it is also possible to evaluate attack classification accuracy in case that the detectors are capable of determining attack types, e.g., by matching them with a list of known and labeled meta-alerts.

\subsubsection{Evaluation of Alert Aggregation Techniques}

Intrusion detection techniques as stated in the previous section often raise large amounts of alerts for some attack steps, where the vast majority of these alerts are duplicates and only have little value to operators that monitor IDSs. Alert aggregation therefore attempts to merge these alerts to reduce the workload of operators and ease the identification of urgent alerts that require immediate actions. On top of that, advanced aggregation techniques are capable of recognizing patterns of alert occurrences and are able to connect attack steps to attack scenarios \cite{landauer2020dealing}. 

In order to merge alerts and attack steps, it is obviously necessary to have datasets at hand that contain repetitions of the same or similar attacks. Unfortunately, these datasets are rare even though they are urgently needed in research \cite{navarro2018systematic}. We therefore propose to forensically analyze our datasets with a desired selection of IDSs to obtain sequences of alerts that are used for aggregation. Similar to the evaluation of IDSs, the variations of our attack scenarios come in handy as they yield different alert patterns for each dataset, e.g., variable amounts of alerts for scans with varying duration or optional alerts caused by commands that the attacker only carries out with certain probabilities. This allows to evaluate whether alerts are indeed aggregated with the same attack types independent of slight variations that occur in real-world environments.

\subsubsection{Evaluation of User Profiling Approaches}

User profiling is a trending research topic that aims to create a profile for each user and then use these profiles to group users by their behavior or role. For this, algorithms based on pattern mining read out access logs that detail all page visits by each user \cite{park2018simpler}. Note that this application scenario is not related to cyber attacks, because only the simulation of normal user behavior is relevant. Due to the fact that our simulated users have specific roles (e.g., WordPress editor or administrator) and visit all pages based on transition probabilities, they clearly follow their own behavior profiles. The main advantage over real data is that it is easy to adjust these profiles according to the respective use case and to quantitatively compare their similarities, which is useful for evaluations and cannot be replicated with humans.

\subsection{Limitations}

Despite all aforementioned benefits of our log dataset, we recognize some limitations. Most important, the user simulation that generates a baseline of normal behavior for our collection of log datasets is obviously limited by the extent of our state machines. On the other hand, real datasets that contain traces of humans interacting with the monitored environments always have the possibility to involve artifacts caused by deliberate or accidental misuse of the systems that could yield incorrect alerts by IDSs. Despite our efforts to generate complex user behavior, we therefore cannot ensure that false positive rates achieved on our datasets are representative for real-world systems. Nonetheless, we are convinced that our synthetic datasets have significant advantages over real ones, as they can be freely published without the need to anonymize artifacts due to privacy concerns and may be arbitrarily recreated in modified use-cases if necessary.

We also point out that we aimed to generate the log data in the most realistic way possible, meaning that we did not configure the logging frameworks to collect data on the highest level of granularity, but instead used standard or default configurations wherever applicable. In case that logging levels need to be adapted, it is always possible to replay the attack scenarios on our open-source testbeds. 

As part of varying the parameters of our testbed when generating TSMs from TIMs, we also decided to leave configurations of logging services unchanged in order to ensure that our labeling rules do not accidentally leave some events unlabeled. We leave the task of extending our labeling rules for this kind of variations for future work.

\section{Conclusion} \label{conclusion}

In this paper we present a collection of eight synthetic log datasets for evaluation of intrusion detection systems. We collect our datasets from testbeds generated by a model-driven methodology for testbed setup and labeling. This enables to repeat the data collection procedure arbitrary many times while at the same time varying several parameters of the simulation with low manual effort. In addition, it is simple to scale the network and extend it with additional components or services. Our datasets are openly accessible and maintainable as all code required to deploy testbeds, run simulations, and assign labels to log events is available open-source. Our datasets thus solve several problems that are prevalent in existing datasets, including control over the simulation parameters, presence of repeated attack executions in similar environments, generation of ground truth tables, complexity of the network, preprocessing of logs to protect sensitive information, and more.

Our log datasets address the common use-case of an attack on the infrastructure of a enterprise IT network. In particular, the attack scenario involves reconnaissance scans, brute-force password cracking, data exfiltration, as well as utilization of various tools and exploits to eventually obtain system access. To generate a realistic baseline of normal behavior, we simulate user activity by extensive state machines that are specifically designed to utilize services such as mail platforms and file shares. We primarily created our dataset to provide diverse attack vectors that challenge many different detection techniques, however, we also foresee applications that go beyond IDS evaluation, in particular, alert aggregation and user profiling. We see our dataset as the first in a series and foresee to extend the labeling rules to more files and attack steps in upcoming versions. For future work, we plan to extend user simulations to run on Windows hosts and mobile devices, and to create testbeds for new use-cases such as Internet-of-Things.

\section*{Acknowledgments}

This work was partly funded by the FFG projects INDICAETING (868306) and DECEPT (873980), and the EU projects GUARD (833456) and PANDORA (SI2.835928).

\bibliographystyle{ieeetr}
\bibliography{bibliography}
	
\end{document}